\documentclass[aps,floats,floatfix,showpacs,preprintnumbers,amssymb,prd,twocolumn,superscriptaddress,nofootinbib,nolongbibliography,reprint]{revtex4-2}
\usepackage{comment}
\usepackage{Journal_names}
\usepackage{booktabs,chemformula}
\usepackage{amssymb}
\usepackage{xspace}
\usepackage{nicefrac}
\usepackage{braket}

\newcommand{\bx}{{\boldsymbol x}}
\newcommand{\by}{{\boldsymbol y}}

\newcommand{\ie}{\emph{i.e.},~}
\newcommand{\eg}{\emph{e.g.},~}

\definecolor{cornellGreen}{HTML}{6EB43F}

\definecolor{darkpurple}{HTML}{6A0DAD}

\definecolor{orangeish}{HTML}{DC6232}

\usepackage{graphicx}
\usepackage{dcolumn}
\usepackage{bm}
\usepackage{ragged2e}

\usepackage[normalem]{ulem}

\usepackage[pdfnewwindow=true,      
   colorlinks=true,    
  linkcolor=blue,     
 citecolor=blue,     
filecolor=blue,  
urlcolor=blue,      
final=true,
]{hyperref}

\hypersetup{
    colorlinks=true,
    citecolor=[rgb]{.1, .7, .6},
    linkcolor=[rgb]{.2, .55, .95},
    filecolor=magenta,
    urlcolor=[rgb]{.1, .7, .6},
}

\begin{document}
\setlength{\abovedisplayskip}{7pt}
\setlength{\belowdisplayskip}{7pt}

\preprint{DESY-26-165}
\title{Self-gravity depletes superradiant clouds}
\author{Mateja Bo\v skovi\'c}
\affiliation{Deutsches Elektronen-Synchrotron DESY, Notkestraße 85, 22607 Hamburg, Germany}

\author{Lucas Gathercole}
\affiliation{Rudolf Peierls Centre for Theoretical Physics, University of Oxford, Parks Road, Oxford OX1 3PU, United Kingdom}

\author{Thomas F.~M.~Spieksma}
\affiliation{School of Natural Sciences, Institute for Advanced Study, 1 Einstein Drive, Princeton, NJ 08540, USA}
\affiliation{Rudolf Peierls Centre for Theoretical Physics, University of Oxford, Parks Road, Oxford OX1 3PU, United Kingdom}

\author{Samuel J.~Witte}
\affiliation{Rudolf Peierls Centre for Theoretical Physics, University of Oxford, Parks Road, Oxford OX1 3PU, United Kingdom}
\affiliation{Deutsches Elektronen-Synchrotron DESY, Notkestraße 85, 22607 Hamburg, Germany}
\affiliation{II. Institute of Theoretical Physics, Universität Hamburg, 22761, Hamburg, Germany}

\begin{abstract}
Superradiant clouds of ultralight scalars around black holes are often modeled as free fields, whose levels grow independently and decay slowly through gravitational-wave emission. We show that this limit is never realized. The cloud's self-gravity redistributes particles between levels, just as in the case of axion self-interactions, and dominates over self-interactions for decay constants $f_a\gtrsim6\times10^{16}\,{\rm GeV}\,(\alpha/0.1)$, with $\alpha$ the gravitational fine-structure constant. Self-gravity depletes the cloud by orders of magnitude and feeds more extended levels, weakening all signals that rely on the existence of a large cloud mass. A binary companion, however, can partly offset the depletion. Any assessment of superradiant signals must therefore account for both self-gravity and the cloud's environment.
\end{abstract}

\maketitle

\noindent{\bf{\em Introduction.}} Ultralight bosons can accumulate around spinning black holes (BHs) through superradiance, which extracts energy and angular momentum from the BH, and builds a hydrogen-like cloud (the so-called ``gravitational atom'') characterized by discrete bound states labeled by an analogous set of quantum numbers $\ket{n\ell m}$~\cite{1971JETPL..14..180Z,1972JETP...35.1085Z,Press:1972zz,Starobinskii:1973vzb,Damour:1976kh,Zouros:1979iw,Detweiler:1980uk,Arvanitaki:2009fg,Arvanitaki:2010sy,East:2018glu,Brito:2015oca}. For a free field, this process is controlled by the gravitational fine-structure constant $\alpha\equiv GM\mu$ and the dimensionless BH spin $\tilde{a}$, where $M$ and $\mu$ are the BH and boson masses.\footnote{We use $c=\hbar=1$ and $G=M_{\rm Pl}^{-2}$.} This leads to a rich array of observational signatures that range from gaps in the BH mass--spin plane~\cite{Arvanitaki:2009fg,Arvanitaki:2010sy,Brito:2014wla,Arvanitaki:2014wva,Arvanitaki:2016qwi,Cardoso:2018tly,Ng:2020ruv,Baryakhtar:2020gao,Khalaf:2024nwc,Witte:2024drg,Caputo:2025oap,Ning:2026ebu}, to continuous gravitational waves (GWs) from the cloud~\cite{Arvanitaki:2014wva,Arvanitaki:2016qwi,Brito:2017wnc,Brito:2017zvb,Tsukada:2018mbp,Palomba:2019vxe,Zhu:2020tht,Tsukada:2020lgt,Baryakhtar:2020gao,LIGOScientific:2021rnv,Kyriazis:2025fis,Ellis:2026gwt,Collaviti:2024mvh}, to imprints on binary inspirals~\cite{Baumann:2018vus,Baumann:2019ztm,Zhang:2019eid,Baumann:2021fkf,Baumann:2022pkl,Tomaselli:2023ysb,Takahashi:2023flk,Duque:2023seg,Brito:2023pyl,Boskovic:2024fga,Tomaselli:2024bdd,Tomaselli:2024dbw,Takahashi:2024fyq,Dyson:2025dlj,Tomaselli:2025jfo,Li:2025ffh,Kim:2025wwj,Boskovic:2025ixx,Xu:2026aic,Boskovic:2026eth,Khalaf:2026jap,Cheng:2026dsq,Xu:2026cky,Keijzer:2026vul,Li:2026foy}, providing a unique opportunity to probe bosons in the $10^{-20}$--$10^{-10}$\,eV range. Axions and axion-like particles are among the best-motivated candidates in this range: they can explain the conservation of charge-parity in QCD~\cite{Peccei:1977hh,Weinberg:1977ma,Wilczek:1977pj}, are compelling dark matter candidates~\cite{Preskill:1982cy,Abbott:1982af,Dine:1982ah,Hui:2016ltb}, and emerge naturally in string theory~\cite{Svrcek:2006yi,Arvanitaki:2009fg}.

With the exception of BH spin-down, these signatures are usually computed for a free field. Each level then grows independently and decays only through GW emission, leaving a cloud of mass $M_{\rm c}\sim\alpha M$ that, for small $\alpha$, can survive for the age of the Universe. For axions,\footnote{Similar statements also apply to low-energy scalars with approximate symmetries protecting their mass.} this assumption is not justified \emph{a priori}. Their periodic potential contains an attractive quartic interaction, $\mathcal L\supset\lambda\phi^4/4!$ with $\lambda \simeq \mu^2/f_a^2$, which drives $2\to2$ transitions between orbital levels, $\ket{\psi_1}\times\ket{\psi_2}\to\ket{\psi_3}\times\ket{\psi_{\rm out}}$, where the fourth leg is an ``off-shell'' forced oscillation that carries a flux out of the cloud: it is either absorbed by the BH or emitted to infinity \cite{Gruzinov:2016hcq,Baryakhtar:2020gao,Omiya:2022gwu,Witte:2024drg,Cannizzaro:2026qgf}. These transitions drain the dominant level and populate higher ones, which stalls the growth of the cloud and the spin-down of the BH. Their rates scale as $1/f_a^{4}$, and are therefore suppressed at large decay constants (although they can still matter for Planck scale $f_a$ \cite{Cannizzaro:2026qgf}).

\begin{figure}
    \centering
    \includegraphics[width=0.99\linewidth]{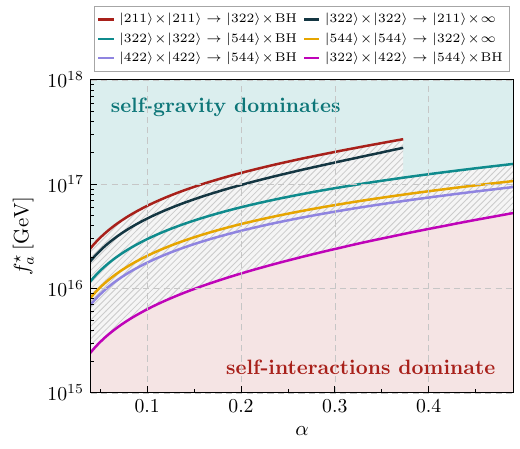}
    \caption{Decay constant $f_a^\star$ at which the SG and SI rates of a $2\to2$ channel are equal, for representative channels at $\tilde a=0.95$.}
    \label{fig:cross-over}
\end{figure}

In this \emph{Letter}, we show that the ``free-field'' (or ``non-interacting'') limit never exists, even for $\lambda\to0$. The cloud's self-gravity (SG), so far considered only for the shift it induces in the level frequencies~\cite{Baryakhtar:2020gao,Siemonsen:2022yyf,Cannizzaro:2023jle,May:2024npn,Kim:2025wwj,Boskovic:2026eth}, acts as a quartic interaction that drives the same $2\to2$ transitions as axion self-interactions (SI), with the same selection rules at leading order. This interaction \emph{does not depend on $f_a$}, and it dominates over SI for decay constants above
\begin{equation}
f_a^\star\simeq \frac{M_{\rm Pl}\,\alpha}{K}\simeq6\times10^{16}\,{\rm GeV}\left(\frac{\alpha}{0.1}\right)\,,
\label{eq:fstar}
\end{equation}
with $K\simeq20$ for the channel that depletes $\ket{211}$ and $K\simeq25$--$60$ for the other channels among $\ket{211}$, $\ket{322}$ and $\ket{544}$ (Fig.~\ref{fig:cross-over}). The QCD axion, for example, lies deep in this regime, with $f_a^\star \simeq 0.01 \, f_a \, (10 \, M_\odot/M)(\alpha/0.1)^2$.

Self-gravity thus sets a floor on the strength of the $2\to2$ interactions, which survives as $f_a$ is pushed toward the Planck scale. As a result, superradiant clouds always decay much faster than a free field, and are both lighter and more spatially extended. This holds even at small $\alpha$, where clouds were thought to live longest. This requires a reassessment of various phenomenological implications of scalar superradiance.

\noindent{\bf{\em Self-gravity as a quartic interaction.}}
We consider a real scalar $\phi$ around a Kerr BH, in the nonrelativistic regime $\alpha\ll1$ and to leading order in $M_{\rm c}/M$. Relativistic corrections are treated in the \emph{Supplemental Material} (SM). Writing $\phi=(\psi e^{-i\mu t}+{\rm c.c.})/\sqrt{2\mu}$ and integrating out the fast oscillations, the cloud dynamics is governed by the Hamiltonian
\begin{equation}
\begin{aligned}
H &=\int_{\bx}\bar\psi\left(-\frac{\nabla^2}{2\mu}-\frac{\alpha}{r}\right)\psi\, + H_{\rm quart}\,, \\ 
H_{\rm quart} &\equiv \frac12\int_{\bx\by}\bar\psi(\bx)\psi(\bx)\,V(\bx,\by)\,\bar\psi(\by)\psi(\by)\,,
\end{aligned}
\end{equation}
where $\int_{\bx}\equiv\int{\rm d}^3x$ and a bar denotes complex conjugation. The quartic interaction sector $H_{\rm quart}$ is described by the kernel $V$, which receives three contributions $V=V_{\rm SI}+ V_{\rm SG} + V_{\rm rec}$. The first two arise from the axion SI and the Newtonian potential of the cloud,
\begin{equation}
V_{\rm SI}=-\frac{\lambda}{8\mu^2}\,\delta^3(\bx-\by)\,,\qquad
V_{\rm SG}=-\frac{G\mu^2}{|\bx-\by|}\, ,
\label{eq:V}
\end{equation}
and the third arises from BH recoil. We must include the BH's own kinetic energy in the Hamiltonian, and in atom-centered coordinates, momentum conservation allows this to be written in terms of the cloud momentum via a quartic kernel (see SM)
\begin{equation}
H_{\rm rec}=\frac{\mathbf P_\psi^2}{2M}\quad \implies \,\,\,
V_{\rm rec}=\frac{\hat{\mathbf p}_{\bx}\!\cdot\hat{\mathbf p}_{\by}}{M}\,,
\label{eq:Vrec}
\end{equation}
where $\mathbf P_\psi=-i\int_{\bx}\bar\psi\nabla\psi$ is the cloud momentum and $\hat{\mathbf p}=-i\nabla$. By the equivalence principle, the same recoil interaction appears in the BH frame as the inertial force associated with its acceleration towards the cloud; the two descriptions yield identical transition amplitudes. 

Expanding $\psi=\sum_a \, c_a \, \psi_a$ in the atomic levels $a=\ket{n\ell m}$ reduces the dynamics to that of the mode amplitudes $c_a$, where $|c_a|^2=N_a$ is the occupation number. They obey $i\dot c_a=\partial H/\partial\bar c_a$, with 
\begin{align}
H&=\sum_a E_a |c_a|^2+\frac12\sum_{abcd}U_{abcd}\,\bar c_a c_b\bar c_c c_d\,,
\label{eq:HC}\\
U_{abcd}&=\int_{\bx\by}\bar\psi_a(\bx)\psi_b(\bx)\,V(\bx, \by)\,\bar\psi_c(\by)\psi_d(\by)\,,
\label{eq:U}
\end{align}
where $E_a$ are the level energies and $U=U^{\rm SI}+U^{\rm SG}+U^{\rm rec}$. The recoil arises at the same order as the Newtonian term, $N_aU^{\rm rec}\sim N_aU^{\rm SG}\sim\mu\alpha^2M_{\rm c}/M$. It contributes only to dipole transitions, where it can reverse the sign of the gravitational amplitude. From now on, SG refers to the sum $U^{\rm SG}+U^{\rm rec}$. The ratio of SI to SG is
\begin{equation}
\frac{U^{\rm SI}}{U^{\rm SG}}=\frac{\alpha^2}{8}\left(\frac{M_{\rm Pl}}{f_a}\right)^2\frac{\hat U^{\rm SI}}{\hat U^{\rm SG}}\,,
\label{eq:ratio}
\end{equation}
where $\hat U$ denotes the matrix elements in Bohr-radius units $r_{\rm B}=1/(\alpha\mu)$. This ratio gives the scaling of Eq.~\eqref{eq:fstar}.

The diagonal terms of Eq.~\eqref{eq:HC} (direct, $U_{aabb}$, and exchange, $U_{abba}$) shift the levels~\cite{Baryakhtar:2020gao,May:2024npn,Witte:2024drg,Boskovic:2026eth,Cannizzaro:2026qgf}. The off-diagonal ones induce the level mixings $\ket{\psi_1}\times\ket{\psi_2}\to\ket{\psi_3}\times\ket{\psi_4}$, which are efficient when three of the modes are highly occupied and the fourth decays rapidly. If the fourth mode is bound (has a frequency below $\mu$), then it is absorbed by the BH. Summing over all such $\psi_4$ gives a total scattering rate to the horizon, which is generally dominated by the $\ell=0$ component.\footnote{In some cases, $\ell=2$ final states that are resonant can dominate over the $\ell=0$ contribution~\cite{Cannizzaro:2026qgf}.} To account for radiation to infinity, we must similarly include the unbound Coulomb waves in the sum over $\psi_4$ ~\cite{Baryakhtar:2020gao,Witte:2024drg}. Calculating the radial flux through a sphere at infinity gives the total scattering rate. The normalized occupations $\varepsilon_a\equiv N_a/(GM_0^2)$ then obey
\begin{equation}
\dot{\varepsilon}_a = \gamma^{\rm SR}_a\, \varepsilon_a + \sum_{bc} \mathcal{S}_{bc}^a \, \gamma_{abc} \, \varepsilon_a \varepsilon_b \varepsilon_c + {\rm GW}\,.
\label{eq:rate}
\end{equation}
Here $M_0$, $\tilde a_0$ and $\alpha_0\equiv GM_0\mu$ are the initial BH mass, spin and coupling, $\gamma^{\rm SR}_a$ is the superradiant rate, $\mathcal{S}^a_{bc}=\pm1$ or $\pm2$ is the number of particles the channel adds to or removes from level $a$, and ``GW'' stands for GW annihilation and level transitions~\cite{Cannizzaro:2026qgf}. The scattering rate of each channel combines SG and SI,
\begin{equation}
\gamma(f_a)=\gamma_{\rm SG}\left[1+\kappa\left(\frac{f_a^\star}{f_a}\right)^{\!2}\,\right]^2,\qquad f_a^\star\equiv M_{\rm Pl}\left(\frac{\gamma_{\rm SI}}{\gamma_{\rm SG}}\right)^{1/4},
\label{eq:gamma}
\end{equation}
with $\gamma_{\rm SG}$ the SG rate, $\gamma_{\rm SI}$ the SI rate at $f_a=M_{\rm Pl}$, and $\kappa=\pm1$ the relative sign of the SG and SI matrix elements for that channel. Equation~\eqref{eq:gamma} defines $f_a^\star$ channel by channel -- Eq.~\eqref{eq:fstar} is its leading-order form. For typical BH final states, $1-|\kappa| $ is typically negligibly small (this follows from the fact that the contribution of $\ell = 2$ states to $\psi_{\rm out}$ is typically very small); for scattering processes with axion emission, however, the sum over $\ell-$multipoles can lead to order-one inference factors.  For the two channels that involve $\ket{211}$, which are dipole transitions, recoil makes $\kappa=-1$: SG and an attractive quartic then interfere destructively, and the rate vanishes at $f_a=f_a^\star$. Each channel has its own value of $f_a^\star$, so there exists no point in parameter space for which all scattering processes are negligible. Other channels have either sign, $\ket{322}\times\ket{322}\to\ket{544}\times{\rm BH}$, for example, has $\kappa=+1$.

We compute $\gamma_{\rm SI}$, $\gamma_{\rm SG}$ and $\kappa$ for every channel by evaluating the kernels of Eqs.~\eqref{eq:V} and~\eqref{eq:Vrec} between Kerr quasi-bound states and propagating the fourth quantum with the Kerr Green's function~\cite{Dolan:2007mj,Witte:2024drg,Cannizzaro:2026qgf}.\footnote{The code is publicly available at \url{https://github.com/SamWitte/Axion_SR}.} Note that this procedure leads to non-negligible $\mathcal{O}(1)$ corrections with respect to the naive non-relativistic expansion, in which one expands in a basis comprised of hydrogenic bound states and the flat space continuum (SM). This discrepancy arises predominantly because the $\ell=m=0$ final states, which typically provide the largest contribution to the bound oscillation, are not suppressed at small radial distances, and therefore have a stronger response to small relativistic effects. In the SM, we show the impact of the leading order PN corrections on various scattering processes, and discuss the procedure for computing fully relativistic rates. Finally, we note that we compute the rates to the BH at $\tilde a=0.95$ and rescale them to lower spins with $\tilde r_+\equiv1+\sqrt{1-\tilde a^2}$; this procedure in general can lead to a mild over-estimation, by a factor of up to $1.5$ after spin-down. 

\begin{figure*}[t!]
    \centering
    \includegraphics[width=0.99\linewidth]{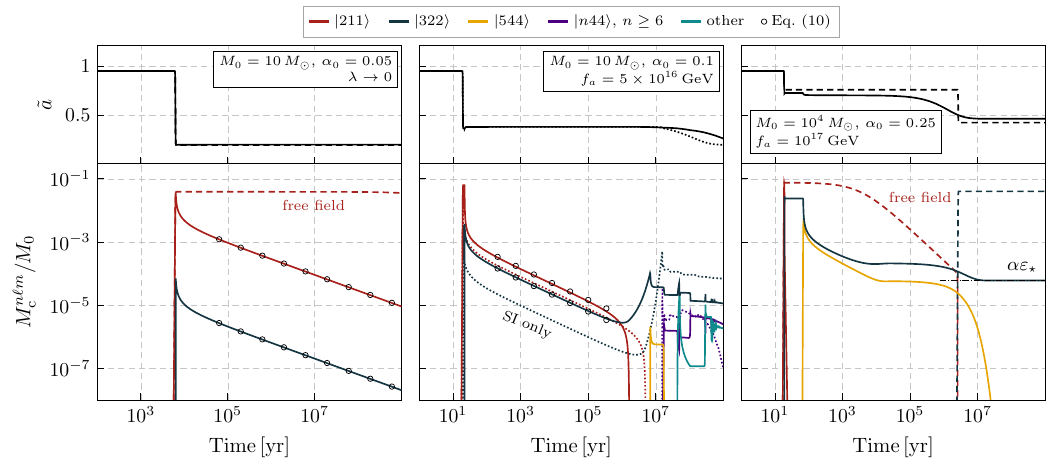}
    \caption{BH spin and cloud mass per level, $M_{\rm c}^{n\ell m}/M_0=\alpha_0\varepsilon_{n\ell m}$, for initial spin $0.95$. Solid lines include SG. \emph{Left:} SG alone against the free field (dashed); circles show Eq.~\eqref{eq:depletion}. \emph{Middle:} SG and SI against SI alone (dotted); the curves for $\ket{n44}$, $n\geq6$, and for the other levels are sums. \emph{Right:} SG and SI against the free field (dashed) for a heavier BH;  the dash-dotted line is the remnant of Eq.~\eqref{eq:relic}.}
    \label{fig:evolution}
\end{figure*}

\noindent{\bf{\em Cloud evolution.}}
For a free field, the fastest-growing level $\ket{a}$ spins the BH down to the threshold of that level, $\tilde a_{\rm sat}\simeq4m_a\alpha/(m_a^2+4\alpha^2)$, reaching an occupation $\varepsilon_a^{\rm sat}\simeq(\tilde a_0-\tilde a_{\rm sat})/m_a$. It then decays slowly through GW annihilation. Once the next superradiant level has grown from its quantum seed, it spins the BH down further and $\ket{a}$ is reabsorbed. The cloud thus passes through a succession of one-level systems~\cite{Arvanitaki:2014wva,Baryakhtar:2020gao}.

Quartic interactions change this picture.  A higher partner level $\ket{b}$ with $m_b=2m_a$ is fed most efficiently by $\ket{a}\times\ket{a}\to\ket{b}\times{\rm BH}$, at a rate $\gamma_{\rm BH}$. The channel $\ket{b}\times\ket{b}\to\ket{a}\times\infty$ and the GW transition $\ket{b}\to\ket{a}\times g$ return particles at rates $\gamma_\infty$ and $\gamma_{\rm tr}$. For sufficiently small $\alpha$, the evolution is captured by the two-state system of $\ket{211}$ and $\ket{322}$: all other channels stay negligible while these levels are populated, and the system can be solved explicitly (SM; see also~\cite{Gruzinov:2016hcq,Baryakhtar:2020gao,Omiya:2022gwu}). Because this simple system is a useful guide, we first describe it in the SG-dominated limit $f_a \gg f_a^\star$, returning later to when and how higher-order states appear. The early evolution of the two-state system proceeds in three phases:

\emph{(i) Saturation.} In general, quartic interactions  are stimulated by the occupation of $\ket{b}$, \ie the rate scales as $N_a^2 (1+N_b)$, with $N_{a,b}$ being the respective occupation numbers. Initially, $N_b$ is negligible, and the rate only becomes efficient when the occupation of $\ket{a}$ becomes extremely large (quantitatively, we find that occupation numbers near saturation are required). Here, $\ket{b}$ cannot grow sufficiently fast, and the first spin-down is essentially unchanged from the free-field limit (SM). Note that for $f_a\ll f_a^\star$, SIs are sufficiently strong to halt the growth prior to spin down~\cite{Baryakhtar:2020gao,Witte:2024drg,Cannizzaro:2026qgf}.

\emph{(ii) Depletion.} After spin down, the ratio $\varepsilon_{322}/\varepsilon_{211}$ quickly relaxes to a constant $r_\star \simeq\gamma_{\rm BH}/(2\gamma_\infty)$ (GW level transitions can be neglected here, SM),  allowing one to directly solve for the evolution of the $\ket{211}$ and $\ket{322}$ states independently:
\begin{equation}
\dot\varepsilon_{211}=-C\,\varepsilon_{211}^3
\quad\Longrightarrow\quad
\varepsilon_{211}(t)\simeq\sqrt{\frac{GM}{2\,c_{\rm SG}\,\alpha^{11}\,t}}\,,
\label{eq:depletion}
\end{equation}
where $C\equiv c_{\rm SG}\,\alpha^{11}/(GM)\simeq3\gamma_{\rm BH}^2/(4\gamma_\infty)$ and $t$ is measured from saturation. Here $\alpha$ and $M$ are the current values, which lie a few percent from $\alpha_0$ and $M_0$ (note this few percent can, in some cases, play a meaningful role due to the strong scaling with $\alpha$). The late-time form applies once $t\gg t_{\rm dep}\equiv[2C(\varepsilon_{211}^{\rm sat})^2]^{-1}$, after which the occupation no longer depends on its value at saturation. For a $10\,M_\odot$ BH with $\alpha=0.1$, $t_{\rm dep}\simeq4$\,days, whereas GW annihilation takes $10^5$--$10^6$\,yr. SG thus depletes the cloud long before GW emission matters -- this is one of the key insights of this work.

\emph{(iii) Remnant.} The depletion stage of $\{\ket{211},\ket{322}\}$  shuts off near $t_\mathrm{end} \simeq 2\gamma_\infty / (3\gamma^\mathrm{SR}_{322} \gamma_\mathrm{BH})$, \ie when the superradiant growth of $\ket{322}$ drives the BH spin below the threshold of $\ket{211}$, causing it to be reabsorbed. 

After the $\ket{211}$ disappears, the $\ket{544}$ appears, generating a new two-state system (comprised of $\{\ket{322},\ket{544} \}$). Here, however, the GW transition is quadrupolar, and is thus competitive with the superradiant and scattering rates; together these processes  cap the growth of $\ket{322}$ below the free-field value. As the $\ket{322}$ superradiance terminates, its occupancy flows towards the fixed point
\begin{equation}
\varepsilon_\star=\frac{\gamma_{\rm tr}}{\gamma_{\rm BH}}\,,
\label{eq:relic}
\end{equation}
while the occupancy of $\ket{544}$ asymptotes to zero. Note however that $\varepsilon^{(n)}_{\star}$ for each pair $\{\ket{322},\ket{n44}\}$ decreases as $n$ grows. Consequently, it becomes favorable for a cascade to occur where the $\ket{n44}$ states are populated on intermediate scales as the occupancy of $\ket{322}$ decreases in the stepwise-like fashion, following the decrease of $\varepsilon^{(n)}_\star$. On even longer timescales, superradiant growth of higher-$n$ states excites, via quartic coupling,  many more states, and the two-state approximation no longer applies.

We test this picture with the numerical setup of~\cite{Witte:2024drg,Cannizzaro:2026qgf}. We evolve the minimal set of quasi-bound states identified in~\cite{Cannizzaro:2026qgf}, including a tower of $\ket{n44}$ states (motivated by the unimpeded decay of the $\ket{322}$ state) -- this amounts to 215 quasi-bound states coupled through $\sim 70,000$ scattering channels, together with the BH mass and spin, and including both SI and SG. SG calculations are only computed up to $n \leq 20$ for simplicity.\footnote{The growth rates of some high-$n$ $\ket{n44}$ states are nearly degenerate, and small differences in the calculation can change which of these states appear. Environmental effects are likely to stabilize the system.} The calculation includes relativistic superradiance and SI scattering rates~\cite{Cannizzaro:2026qgf}. Figure~\ref{fig:evolution} shows three cases, all with initial spin $\tilde a_0=0.95$.

With SG alone (left panel, $\alpha_0=0.05$), the cloud saturates as in the free field scenario, and then follows Eq.~\eqref{eq:depletion}, with $\varepsilon_{211}\simeq5.7\,(t/{\rm yr})^{-1/2}$ and $\varepsilon_{322}/\varepsilon_{211}=2.2\times10^{-3}$ (the two-level model reproduces both to within $1\%$). The cloud is several thousand times less massive than the free field case by the end of the evolution.

The middle panel adds SI ($\alpha_0=0.1$, $f_a=5\times10^{16}$\,GeV). The three stages are unchanged, with $C$ and $r_\star$ now depending on $f_a$. This $f_a$ lies close to $f_a^\star$ of the return channel, where SG and SI interfere \emph{destructively}, so neither can be neglected: the cloud still decays as $t^{-1/2}$, but it is about twice as heavy as with SI alone (dotted), and somewhat lighter than with SG alone. After $\ket{211}$ is reabsorbed, $\ket{322}$ feeds $\ket{544}$ and a tower of $\ket{n44}$ overtones in a burst-like cadence. With SI alone, this cascade climbs to $n\simeq16$ (see also~\cite{Cannizzaro:2026qgf}); SG cuts it off near $n\simeq10$ and populates levels with $m=5$, $6$ and $8$ instead. At late times, $\varepsilon_{322}$ lies several times below its value for SI alone.

Finally, the right panel shows a heavier BH at larger $\alpha$ ($M_0=10^4\,M_\odot$, $\alpha_0=0.25$, $f_a=10^{17}$\,GeV). Scattering now outruns the response of the BH. The absorbed particles carry no angular momentum, so they raise the BH mass and lower its dimensionless spin below the threshold of $\ket{211}$, which is reabsorbed within $6$\,yr of saturation. The BH is left heavier than in the free-field case, and its spin plateau therefore lies marginally lower. Level $\ket{322}$ inherits $\varepsilon_{322}\simeq0.1$, drops as $\ket{544}$ grows, and then falls from $8\times10^{-4}$ while it spins the BH down between $10^5$ and $10^7$\,yr. Once the spin reaches the threshold of $\ket{322}$, $\ket{544}$ empties, and $\ket{322}$ settles at $\varepsilon_\star=2.4\times10^{-4}$, within $1\%$ of Eq.~\eqref{eq:relic} and far below the free-field value.

In all three cases the dense, long-lived cloud of the free field is absent. Further cases are shown in the SM.\\

\noindent{\bf{\em A binary companion can halt the depletion.}}
As shown in Fig.~\ref{fig:evolution}, scattering populates a plethora of high-$n$ states, thus extending the effective size of the cloud since $r_{\rm c}\sim n^{2} r_{\rm B} $. Consequently, the interaction of the cloud with the BH's astrophysical environment needs to be considered. While we leave a full treatment to~\cite{us}, we provide a proof of principle here. Considering again the two-level system $\{\ket{211},\ket{322}\}$, we add a companion of mass $qM$, with $q\simeq0.1$, on a circular equatorial orbit that co- or counter-rotates with the cloud. The orbit adds its mean anomaly $\vartheta$ and frequency $\Omega$ to the phase space, and the Hamiltonian gains a Keplerian term and a tidal coupling (see~\cite{Boskovic:2025ixx} and SM). The tidal coupling generates, \emph{inter alia}, a new network of bound-bound couplings that can be resonantly enhanced~\cite{Baumann:2018vus,Baumann:2019ztm,Boskovic:2024fga,Tomaselli:2024bdd,Tomaselli:2024dbw,Boskovic:2025ixx}, together with bound-to-unbound transitions (ionization)~\cite{Baumann:2021fkf,Baumann:2022pkl,Tomaselli:2023ysb,Tomaselli:2024bdd}. We start the binary in the early inspiral, where ionization is weak and only a few resonances are crossed, all of them on co-rotating orbits.

\begin{figure}
    \centering
    \includegraphics[width=0.99\linewidth]{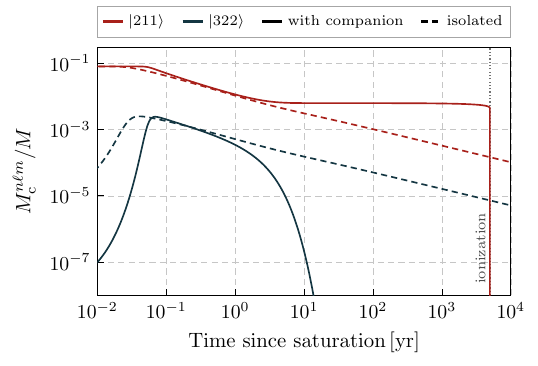}
    \caption{Cloud mass in $\ket{211}$ and $\ket{322}$ against time since saturation, around an isolated BH (dashed) and with a co-rotating companion (solid), for  $\tilde a_0=0.95$ and $M=100\,M_\odot$, $\alpha=0.12$, $q=0.1$ (after $\ket{211}$ saturation). The cloud forms $5\times10^3$\,yr before merger. The companion removes $\ket{322}$ within about $10$\,yr, after which $\ket{211}$ stays nearly constant until the companion ionizes it (dotted line).}
    \label{fig:binary}
\end{figure}

Figure~\ref{fig:binary} shows an example for $M=100\,M_\odot$, $\alpha=0.12$ (after $\ket{211}$ saturation) and a co-rotating orbit. The companion first induces the off-resonant mixing of $\ket{322}$ with $\ket{300}$ over a wide range of orbital frequencies~\cite{Tong:2022bbl,Boskovic:2025ixx}, and the BH absorbs the $\ket{300}$ state. This drains the $\ket{322}$ state, which weakens the stimulated process $\ket{211}\times\ket{211}\to\ket{322}\times{\rm BH}$, and leaves $\ket{211}$ about $30$ times heavier than for the case of an isolated BH. Depending on the cloud formation time, this factor is typically $\mathcal{O}(10)$, with larger values favoring co-rotating orbits (see SM for details). The cloud stays in this state until the companion ionizes it.

By the time of ionization, the binary has entered the LISA band ($10^{-3}$--$10^{-1}$\,Hz) where the large finite-size effects of the cloud~\cite{Baumann:2018vus,Baumann:2019ztm,Boskovic:2026eth} could provide a fingerprint of its presence. The annihilation signal, at $\mu/\pi\simeq77$\,Hz, falls in the band of ground-based detectors, so such a system has signals in both bands.

At larger $\alpha$, where the levels sit at the fixed point, the effect of a companion is more involved. We expect it to limit the cascade to high $n$: those levels extend much closer to the companion and feel a stronger tide early in the inspiral~\cite{us}.

\noindent{\bf{\em Observational consequences and conclusions.}}
For small $\alpha$, the fastest-growing level grows a cloud to its free-field mass, and SG then drains it (unless a companion intervenes). The depletion is so rapid that a cloud observed today is almost certainly not the fully grown, saturated one. With SG alone, Eq.~\eqref{eq:depletion} gives
\begin{equation}
\begin{aligned}
\frac{M_{\rm c}}{M}&\simeq 3\times10^{-5}\left(\frac{5.2}{c_{\rm SG}}\right)^{1/2}\left(\frac{0.05}{\alpha}\right)^{9/2}\\
&\times\left(\frac{M}{10\,M_\odot}\right)^{1/2}\left(\frac{10^8\,{\rm yr}}{t}\right)^{1/2}\,,
\label{eq:Mc}
\end{aligned}
\end{equation}
with $c_{\rm SG}$ as listed in the SM (Table~\ref{tab:cSG}). For reference, a free field would give $M_{\rm c}/M\simeq0.04$ at $\alpha=0.05$. At fixed $\alpha$ all rates scale as $1/M$, so clouds around heavy BHs deplete more slowly: for $M=10^6\,M_\odot$, $\alpha=0.07$ and $t=10^9$\,yr, Eq.~\eqref{eq:Mc} gives $M_{\rm c}/M\simeq5\times10^{-4}$, still $\sim 100$ times below the free-field value.

Equation~\eqref{eq:Mc} holds until $\ket{322}$ spins the BH down further, and $\ket{211}$ is reabsorbed. For a $10\,M_\odot$ BH, this takes $\sim10^7$\,yr at $\alpha_0=0.1$, and more than $10^9$\,yr for $\alpha_0\leq0.075$. For $\alpha_0\gtrsim0.15$, $\ket{211}$ is reabsorbed right after saturation and Eq.~\eqref{eq:Mc} never applies. The cloud then consists of $\ket{322}$, and a series of more weakly bound states. With SG alone, it holds up to $0.7\%$ of the BH mass (at $\alpha_0=0.25$) when $\ket{322}$ takes over, and ends at $M_{\rm c}/M\simeq\alpha\,\varepsilon_\star\sim10^{-5}$--$10^{-4}$ [Eq.~\eqref{eq:relic}]. SI change the depletion coefficient (SM). They are negligible for $f_a\gg f_a^\star$, as for the QCD axion. They make the cloud lighter for $f_a\ll f_a^\star$, and can lead to order one differences near $f_a^\star$. SG therefore affects nearly every signal of the cloud.

\emph{Gravitational waves.} The continuous-wave strain from annihilation scales with the mass of the emitting level. For a $10\,M_\odot$ BH with $\alpha_0=0.1$ and SG alone, the mass in $\ket{211}$ is $900$ times smaller than for a free field only $10^4$\,yr after saturation, and the strain is proportionally weaker. The energy radiated by such a source, which enters the stochastic background~\cite{Brito:2017zvb,Tsukada:2020lgt}, is lower by a factor of $10^6$ over $10^9$\,yr for $\alpha_0=0.05$--$0.1$. Searches and forecasts that assume a free field overestimate the signal accordingly.

\emph{Astrophysical environment.} The backreaction of the cloud on the orbit is proportional to $M_\mathrm{c}$ (SM), so a depleted cloud weakens some of its distinctive signatures, such as floating and sinking orbits~\cite{Baumann:2019ztm,Boskovic:2024fga,Tomaselli:2024bdd,Tomaselli:2024dbw,Boskovic:2025ixx} or ionization-driven dynamical friction~\cite{Baumann:2021fkf,Baumann:2022pkl,Tomaselli:2023ysb,Tomaselli:2024bdd,Tomaselli:2025jfo}. Two effects work against this. First, as shown above for small $\alpha$, the companion can itself halt the depletion. Second, SG moves the cloud to more extended levels, which the companion reaches earlier in the inspiral. Other environments could halt the depletion as well, especially around heavier BHs, \eg star clusters embedded in the cloud~\cite{Cheng:2026dsq}. Baryonic accretion, on the other hand, is not expected to restore the cloud: the gas spins the BH up and feeds $\ket{211}$, but scattering balances this at an occupation far below saturation, in contrast to the growth found for a free field~\cite{Brito:2014wla,Hui:2022sri}.

\emph{Scalar radiation.} One third of the particles lost by $\ket{211}$, about $1$--$2\%$ of the BH mass for $\alpha_0=0.05$--$0.1$, is radiated as non-relativistic scalar waves as a narrow line just above $\mu$. If the scalar couples to matter, this radiation could be searched for directly~\cite{Gavilan-Martin:2026zzw}.

\emph{Black-hole spins.} Spin-down is the least affected. SG leaves the first spin-down as it is for a free field, so it does not change constraints based on $\ket{211}$ alone. Later levels start from a macroscopic seed, so they begin to extract spin tens to hundreds of times earlier, but they do so gradually, and most of the spin is lost by about the same time. Once spin-down is complete, the BH is heavier than for a free field, since it reabsorbs most of the cloud, and its final spin is slightly higher, because the threshold spin grows with $\alpha\propto M$. It is worth noting that current spin-down constraints extend into the regime $f_a \ll f_a^\star$~\cite{Witte:2024drg,Caputo:2025oap}, and thus SG is not expected to alter these results.

It is known that SI can strongly deplete superradiant clouds, and spread energy and angular momentum across a wide range of dynamical scales, even for relatively large decay constants~\cite{Baryakhtar:2020gao,Witte:2024drg,Cannizzaro:2026qgf}. SG is more general: it acts on every scalar, whatever its potential, and couples the levels even around an isolated BH, reshaping the cloud's evolution and morphology. We expect the same for vector clouds~\cite{Baryakhtar:2017ngi,East:2017ovw,East:2017mrj}, since even a free Proca field acquires quartic interactions through its own gravity. The free-field limit is thus \emph{never recovered} and the dense, long-lived cloud on which most signatures of superradiance rely does not survive.

\noindent{\bf{\em Declaration.}}
While completing this work, we became aware of Ref.~\cite{Foster:2026nuz}, which studies self-gravity mediated $2\rightarrow 2$ scattering in the Newtonian limit, and obtains results broadly consistent with ours.

\noindent{\bf{\em Acknowledgments.}}
SJW and TS acknowledge support from a Royal Society University Research Fellowship (URF-R1-231065). This work is also supported by the Deutsche Forschungsgemeinschaft under Germany’s Excellence Strategy -- EXC 2121 ``Quantum Universe'' -- 390833306. TS acknowledges support from the Friends of the Institute for Advanced Study Fund and the Simons Foundation through the Black Holes and Strong Gravity program through Award No.~SFI-MPS-BH-00012593-10. LG acknowledges support from the Gould-Watson Scholarship, and the Science and Technology Facilities Council.

\bibliographystyle{apsrev4-2}
\bibliography{refs}
\appendix

\clearpage

\onecolumngrid
\begin{center}
  \textbf{\large Supplemental Material for Self-gravity depletes superradiant clouds}\\[.2cm]
  \vspace{0.05in}
  { Mateja Bo\v skovi\'c, Lucas Gathercole, Thomas F.~M.~Spieksma, and Samuel J.~Witte}\vspace{.4in}
\end{center}

\twocolumngrid

\setcounter{equation}{0}
\setcounter{figure}{0}
\setcounter{table}{0}
\setcounter{section}{0}
\setcounter{page}{1}
\renewcommand{\theequation}{S\arabic{equation}}
\renewcommand{\thefigure}{S\arabic{figure}}
\renewcommand{\thetable}{S\arabic{table}}
\renewcommand{\thesubsection}{{S.\arabic{subsection}}}
\subsection{Relativistic corrections to quartic interactions}
The main text treats SG through the Newtonian potential of the cloud. Here we set up the fully relativistic calculation of the same $2\to2$ scattering (working in the BH-centered frame, in contrast with the main text), and then show how the Newtonian limit of the main text and the leading post-Newtonian (PN) corrections follow from it.

A real scalar that is minimally coupled to gravity, with the periodic potential $\mathcal V(\phi)=\mu^2f_a^2\,[1-\cos(\phi/f_a)]$, obeys
\begin{align}
\Box_g\phi &= \mathcal V'(\phi)=\mu^2\phi-\frac{\lambda}{6}\phi^3+\dots\,, \label{eq:full}\\
G_{\mu\nu}[g]&=8\pi G\,T_{\mu\nu}[\phi]\,,
\end{align}
with $\lambda=\mu^2/f_a^2$ as in the main text. We write the metric as $g=g_{\rm Kerr}+h$, where $h$ is the perturbation produced by the cloud, and the field as
\begin{equation}
\phi=\sum_a c_a\psi_a + \delta\psi \,+\, {\rm c.c.}
\end{equation}
Here $\psi_a$ are the quasi-bound states of Kerr and $\delta\psi$ is the response of the field to the interactions. Unlike in the main text, the $\psi_a$ are relativistically normalised and carry their full time dependence, so no factor $1/\sqrt{2\mu}$ appears. We work to leading order in $h$ and $\delta\psi$, which requires $M_{\rm c}\ll M$, and assume that the amplitudes $c_a$ vary slowly. The response then obeys
\begin{equation}
    \left(\Box_{\rm Kerr}-\mu^2\right)\,\delta\psi = S\,, \label{eq:master}
\end{equation}
where the source $S$ collects the SI and SG of the occupied levels. For the process $\ket{\psi_a}\times \ket{\psi_b}\to \ket{\psi_c}\times \ket{\psi_{\rm out}}$, the relevant part of the source is
\begin{equation}
    S = \sum_{abc}  S_{abc}\,c_ac_b \bar c_c\,.
\end{equation}
Both orderings of $a$ and $b$ appear in the sum, so a process with $a\neq b$ has the amplitude $S_{abc}+S_{bac}$. For SI the components follow directly from the cubic term of Eq.~\eqref{eq:full},
\begin{equation}
    S_{abc} = -\tfrac{1}{2}\lambda\, \psi_a\psi_b \bar\psi_c\,.
\end{equation}
For SG they require the metric perturbation, which we derive next.
\subsubsection{Gravity of a real scalar} 
\label{sec:relativistic}
The metric perturbation is sourced by pairs of levels, because the stress tensor is quadratic in the field. The pairs of a level and a conjugate level, $\propto c_a\bar c_b$, contribute
\begin{equation}
    \begin{aligned}
        T^{(a \bar b)}_{\mu\nu} &= 2\nabla_{(\mu} \psi_a \nabla_{\nu)}\bar\psi_b\\
        & \quad -g_{\mu\nu}\big(\nabla\psi_a\!\cdot\!\nabla\bar\psi_b+\mu^2\psi_a\bar\psi_b\big)\,.
    \end{aligned}
\end{equation}
These vary slowly, at the difference of two level frequencies, and dominate. Because the field is real, the stress tensor also contains pairs of two levels, $\propto c_ac_b$,
\begin{equation}
    \begin{aligned}
        T^{(a b)}_{\mu\nu} &= 2\nabla_{(\mu} \psi_a \nabla_{\nu)} \psi_b\\
        & \quad -g_{\mu\nu}\big(\nabla\psi_a\!\cdot\!\nabla\psi_b+\mu^2\psi_a\psi_b\big)\,,
    \end{aligned}
\end{equation}
and their conjugates $T^{(\bar a \bar b)}_{\mu\nu}$. These oscillate at about $2\mu$ and describe the exchange of a graviton in the $s$-channel (see also~\cite{Salehian:2020bon}). For each contribution we solve the linearised Einstein equations,
\begin{equation}
    \delta G_{\mu\nu}\left[h^{(ab)}\right] = 8\pi G\, T^{(ab)}_{\mu\nu}\,, \label{eq:lin}
\end{equation}
and likewise for $h^{(a\bar b)}$. To first order, a metric perturbation $h$ changes the d'Alembertian by
\begin{equation}
\begin{aligned}
    \delta\Box &\equiv \Box_g - \Box_{\rm Kerr} \\
    &= -h^{\mu\nu} \nabla_\mu \nabla_\nu -(\nabla_\mu h^{\mu\nu} - \tfrac12\nabla^\nu h)\nabla_\nu\,, \label{eq:dbox}
\end{aligned}
\end{equation}
and we write $\delta\Box^{(ab)}$ for this operator evaluated on $h^{(ab)}$. The SG part of the source is then
\begin{equation}
    S_{abc} =-\tfrac12\left( \delta \Box^{(b\bar c)} \psi_a + \delta\Box^{(a \bar c)}\psi_b + \delta \Box^{(ab)} \bar\psi_c\right).
\label{eq:S}
\end{equation}
In the first two terms, the gravitational field of one incoming level and the outgoing level acts on the other incoming level. The third is the $s$-channel term.

From here on SI and SG are treated in the same way. We solve Eq.~\eqref{eq:master} with the retarded Green's function of Kerr~\cite{Dolan:2007mj,Witte:2024drg,Cannizzaro:2026qgf}. The scattering rate follows from the flux of the number current of $\delta\psi$, $i\,\delta\psi\,\nabla_\mu\overline{\delta\psi}+{\rm c.c.}$, through the horizon and through a sphere at infinity, which give the rates to the BH and to infinity that enter Eq.~\eqref{eq:rate}. Equivalently, $\delta\psi$ can be inserted back into the equations for $c_a$, which gives the change in the occupations directly.
\subsubsection{Relativistic recoil}
\label{sec:rel-recoil}
As in the Newtonian description of the main text, the motion of the BH must be taken into account. In principle Eq.~\eqref{eq:lin} contains it, but only if the gauge is chosen with care. The natural choice, Lorenz gauge in an asymptotically inertial frame, has the BH oscillate about the origin~\cite{Detweiler:2003ci}: by momentum conservation it balances the dipole of the cloud, $\sim r_{\rm c}M_{\rm c}$, and is displaced by $d\sim r_{\rm c}M_{\rm c}/M$. In units of the horizon radius,
\begin{equation}
    \frac{d}{r_+} \sim \frac{n^2}{\alpha^2}\, \frac{M_{\rm c}}{M}\,,
\end{equation}
which can be comparable to or much larger than one. The metric then differs strongly from a Kerr BH at the origin, and the expansion in $h$ and $\delta\psi$ fails. We therefore work in a frame that follows the BH, so the horizon stays in place and the expansion remains under control. 
To find the BH frame exactly means satisfying two `pinning' conditions. First, the radial oscillations of the horizon $\delta r_+ \propto \Sigma h^{rr}$ must have no dipole component, which would describe an overall displacement
\begin{equation}
    0=\oint_{r_+} \Sigma h^{rr}\, \hat \bx\, d\Omega.
\end{equation}
Secondly, we require that the horizon does not rotate. We express the tangential velocity of the horizon as $\mathbf{h}_\tau \equiv h_{\tau i} \mathbf{e}_i$ with respect to some horizon-penetrating time-coordinate $\tau$, so that we can write
\begin{equation}
    0 = \oint_{r_+} \mathbf{h}_\tau \times \hat \bx \, d\Omega,
\end{equation}
to enforce no overall rotation.
Finding a gauge fixing that enforces both conditions is numerically subtle, but can be achieved with a method of Lagrange multipliers, which we do not cover here.
Because this is an accelerated frame, the metric gains non-inertial corrections. Their leading term is 
\begin{equation}
    \delta h_{tt} \simeq -2\,\mathbf{a}_{\rm BH}\cdot\bx\,, 
    \label{eq:rel-recoil} 
\end{equation}
with $\mathbf{a}_{\rm BH}$ the acceleration of the BH. This is the relativistic form of the recoil. Its Newtonian limit is derived below, and Sec.~\ref{sec:recoil} obtains the same interaction in the inertial frame. Notably, phase-dependent matrix elements do depend on the frame (BH vs. atom), but scattering rates do not. They come from squared amplitudes and scale with occupation numbers only, which are frame-independent, so the end results of the two calculations must agree.

\subsubsection{Newtonian limit}
The Newtonian limit follows by expanding in the velocity, $v^2\sim GM/r\sim\alpha^2/n^2$, and keeping the leading term. Then $\partial_t\psi\simeq-i\mu\psi$, and the stress tensor is dominated by the Newtonian density,
\begin{equation}
    T^{(a\bar b)}_{00} \simeq 2\mu^2\psi_a\bar\psi_b \equiv \rho^{(ab)}_{\rm N}\,.
\end{equation}
The metric perturbation reduces to the Newtonian potential, $h_{00}=-2\Phi$, which acts on the field as
\begin{equation}
    \delta\Box\,\psi\simeq -h^{00}\partial_t^2 \psi= -2\mu^2\Phi\,\psi\,.
\end{equation}
Each pair of levels sources its own potential,
\begin{equation}
    \Phi^{(ab)}_{\rm N}(\bx)=-G\!\int_{\by}\frac{\rho^{(ab)}_{\rm N}(\by)}{|\bx-\by|}\,, \label{eq:newton}
\end{equation}
and the SG source of Eq.~\eqref{eq:S} becomes, at leading order in $v$,
\begin{equation}
    S_{abc} \simeq \mu^2\left(\Phi^{(bc)}_{\rm N}\psi_a+\Phi^{(ac)}_{\rm N}\psi_b\right)\,.
\end{equation}
Recoil adds $\mathbf{a}_{\rm BH}\cdot \bx$ to the potential, from $\delta h_{tt}$ above. In the Newtonian limit the cloud accelerates the BH as it would any mass,
\begin{equation}
    \mathbf{a}^{(ab)}_{\rm N}=G\!\int_{\by} \hat\by\, \frac{\rho^{(ab)}_{\rm N}(\by)}{|\by|^2}=-\nabla\Phi^{(ab)}_{\rm N}(0)\,,
\end{equation}
so the potential in the frame of the BH is
\begin{equation}
\begin{aligned}
    \Phi^{(ab)}_{\rm N}+\mathbf a^{(ab)}_{\rm N}\!\cdot\bx &= -G\!\int_{\by}\,\rho^{(ab)}_{\rm N}(\by) \\
    &\quad\times\left(\frac{1}{|\bx-\by|}-\frac{\bx\cdot\by}{|\by|^3}\right)\,.
\end{aligned}
\end{equation}
The same form appears in Ref.~\cite{Kim:2025wwj}, which works in the BH-centered frame at the Newtonian level.
\subsubsection{Post-Newtonian corrections}
We now compute the leading PN corrections, which are of relative order $v^2 \sim \alpha^2/ n^2$ (see, \eg Ref.~\cite{Porto:2016pyg}). This requires the subleading part of $T_{00}$, as well as $T_{0i}\sim\rho v$ and $T_{ij}\sim\rho v^2$. We take them in turn, first in the inertial frame and then in the frame of the BH.

In harmonic gauge, $\partial_\mu\bar h^{\mu\nu}=0$, the second term of Eq.~\eqref{eq:dbox} vanishes, and we write
\begin{equation}
    h_{00} = -2\Phi\,, \qquad
    h_{0i} = -4A_i\,, \qquad
    h_{ij} = -2\Phi\,\delta_{ij}\,,
\end{equation}
The spatial part $h_{ij}$ is needed only at leading order, because it multiplies $\partial_i\partial_j\psi=\mathcal O(v^2)$. The potentials obey
\begin{align}
    \left(\nabla^2-\partial_t^2\right)\Phi &= 4\pi G\left(T_{00}+T_{kk}\right)\,, \\
    \nabla^2 A_i &= 4\pi G\,T_{0i}\,,
\end{align}
so only two combinations of the stress tensor enter. For a pair of levels they are
\begin{align}
    \rho^{(ab)} &\equiv T^{(a\bar b)}_{00}+T^{(a\bar b)}_{kk} = \left(\frac{2\omega_a\omega_b}{\mu^2}-1\right)\rho^{(ab)}_{\rm N}\,, \label{eq:rhoPN}\\
    j^{(ab)}_i &\equiv -T^{(a\bar b)}_{0i} = i\left(\omega_a\psi_a\partial_i\bar\psi_b-\omega_b\bar\psi_b\partial_i\psi_a\right)\,,
\end{align}
where $\omega_a$ is the frequency of level $\ket{a}$, and the density $\rho^{(ab)}$ reduces to the Newtonian one at leading order.  On these potentials, Eq.~\eqref{eq:dbox} gives
\begin{equation}
    \delta\Box\,\psi = 2\Phi\left(\partial_t^2+\nabla^2\right)\psi - 8A_i\,\partial_t\partial_i\psi\,.
    \label{eq:dboxPN}
\end{equation}
The source keeps the form of Eq.~\eqref{eq:S}, with $\delta\Box$ evaluated on these potentials. For the slowly varying pairs, which oscillate at $\omega_{ab}\equiv\omega_a-\omega_b$,
\begin{align}
    \left(\nabla^2+\omega_{ab}^2\right)\Phi^{(ab)} &= 4\pi G\,\rho^{(ab)}\,, \\
    \nabla^2 A^{(ab)}_i &= -4\pi G\, j^{(ab)}_i\,.
\end{align}
Across the cloud $\omega_{ab}r\sim\alpha$, so the retardation can be expanded to leading order. Since $\rho^{(ab)}$ is a multiple of the Newtonian density [Eq.~\eqref{eq:rhoPN}],
\begin{align}
    \Phi^{(ab)} &= \left(\frac{2\omega_a\omega_b}{\mu^2}-1\right)\Phi^{(ab)}_{\rm N}+\Phi^{(ab)}_{\rm ret}\,, \\
    \Phi^{(ab)}_{\rm ret}(\bx) &= \frac{G\omega_{ab}^2}{2}\!\int_{\by}\rho^{(ab)}_{\rm N}(\by)\,|\bx-\by|\,, \\
    \mathbf A^{(ab)}(\bx) &= G\!\int_{\by}\frac{\mathbf j^{(ab)}(\by)}{|\bx-\by|}\,,
\end{align}
where $\Phi_{\rm N}$ is the Newtonian potential of Eq.~\eqref{eq:newton}. 

The BH potential, $\Phi_{\rm BH}=-GM/r$, also enters at this order, through terms $\sim\Phi_{\rm BH}\Phi$. They take their simplest form in harmonic coordinates, $\Box_g x^\mu = 0$. With $h$ in harmonic gauge, this only requires a shift of the radial coordinate, $r = r_{\rm BL} - GM$. To the order needed, the potentials then add in an exponent,
\begin{equation}
    g_{00} = -e^{2(\Phi_{\rm BH}+\Phi)}\,, \quad
    g_{0i} = -4A_i\,, \quad
    g_{ij} = e^{-2(\Phi_{\rm BH}+\Phi)}\delta_{ij}\,,
\end{equation}
where each potential is sourced by $T^{00}+T^{kk}$ through the flat-space operator. The BH potential then enters in three ways:
\begin{itemize}
    \item The BH redshifts the potential of the cloud as it acts on $\psi$, $\delta g^{00}=2e^{-2\Phi_{\rm BH}}\Phi$.
    \item The BH blueshifts the energy of the cloud. In the local static frame the flat-space result holds with $\omega\to e^{-\Phi_{\rm BH}}\omega$, and $T^{\hat0\hat0}\simeq2\mu^2\psi_a\bar\psi_b$ dominates. Converting to coordinate components, $T^{00}+T^{kk}=e^{-2\Phi_{\rm BH}}T^{\hat0\hat0}+e^{2\Phi_{\rm BH}}T^{\hat k\hat k}$, gives
    \begin{equation}
        \rho^{(ab)} = \left(\frac{2\omega_a\omega_b}{\mu^2}-1-6\Phi_{\rm BH}\right)\rho^{(ab)}_{\rm N}\,,
    \end{equation}
    which adds to $\Phi$ the potential
    \begin{equation}
        \Phi^{(ab)}_{G}(\bx)\equiv6G\!\int_{\by}\frac{\Phi_{\rm BH}(\by)\,\rho^{(ab)}_{\rm N}(\by)}{|\bx-\by|}\,.
    \end{equation}
    \item The cloud redshifts the mass of the BH, $T^{00}=Mu^0/\sqrt{-g}=Me^{\Phi(0)}$, so $\Phi_{\rm BH}\to\Phi_{\rm BH}\,[1+\Phi(0)]$.
\end{itemize}
In the inertial frame the BH sits at $\mathbf X(t)$ and moves according to
\begin{equation}
    \ddot{\mathbf X}=\mathbf a_{\rm BH}\equiv-\nabla\Phi(0)+4\dot{\mathbf A}(0)\,,
\end{equation}
which is the geodesic equation for a body at rest. The BH's own field, evaluated at the BH, does not act on it, so $\Phi_{\rm BH}$ enters $\mathbf a_{\rm BH}$ only through the density of the cloud, via $\mathbf a_G$ below. Following the split of $\Phi$,
\begin{equation}
\begin{aligned}
    \mathbf a^{(ab)}_{\rm BH} &= \left(\frac{2\omega_a\omega_b}{\mu^2}-1\right)\mathbf a^{(ab)}_{\rm N}+\mathbf a^{(ab)}_G \\
    &\quad +\mathbf a^{(ab)}_{\rm ret}+\mathbf a^{(ab)}_{\rm GM}\,,
\end{aligned}
\end{equation}
with $\mathbf a_{\rm N}\equiv-\nabla\Phi_{\rm N}(0)$, likewise for $\mathbf a_G$ and $\mathbf a_{\rm ret}$, and $\mathbf a_{\rm GM}\equiv4\dot{\mathbf A}(0)$.

We move to the BH frame with $\bx\to\bx-\mathbf X$ and $t\to t-\dot{\mathbf X}\cdot\bx$. To linear order in the cloud, this returns the BH to rest at the origin, leaves $\Phi$ and $\mathbf A$ unchanged, and multiplies $g_{00}$ by $e^{2\mathbf a_{\rm BH}\cdot\bx}$. At leading order this is the recoil term of Eq.~\eqref{eq:rel-recoil}, $\delta h_{00}=-2\,\mathbf a_{\rm BH}\cdot\bx$. The BH's own potential is also retarded. Since the BH accelerates, its expansion contains $-\tfrac12GM\,\partial_t^2|\bx-\mathbf X|=\tfrac12GM\,\mathbf a_{\rm BH}\cdot\hat\bx$, so in the BH frame $\Phi_{\rm BH}\to\Phi_{\rm BH}\big(1-\tfrac12\mathbf a_{\rm BH}\cdot\bx\big)$. The metric in this frame is not in harmonic gauge, so the second term of Eq.~\eqref{eq:dbox} now contributes, and
\begin{equation}
\begin{aligned}
    \delta\Box\,\psi &= 2\Big[(1-2\Phi_{\rm BH})\left(\Phi+\mathbf a_{\rm BH}\!\cdot\bx\right)-\tfrac12\Phi_{\rm BH}\,\mathbf a_{\rm BH}\!\cdot\bx \\
    &\qquad +\Phi_{\rm BH}\Phi(0)\Big]\partial_t^2\psi \\
    &\quad + 2\Phi\nabla^2\psi - 8A_i\,\partial_t\partial_i\psi \\
    &\quad + \mathbf a_{\rm BH}\!\cdot\nabla\psi + \left(\dot{\mathbf a}_{\rm BH}\!\cdot\bx\right)\partial_t\psi\,.
\end{aligned}
\end{equation}

Finally, we must include the $s$-channel pairs, which oscillate at $\omega_a+\omega_b\simeq2\mu$. This high frequency dominates over the gradients in the linearized Einstein equations, $\Box\bar h_{\mu\nu}=-16\pi G\,T_{\mu\nu}$, so the solution is pointwise, $4\mu^2\bar h^{(ab)}_{\mu\nu}\simeq-16\pi G\,T^{(ab)}_{\mu\nu}$. At leading order, the only component of the $s$-channel stress tensor is $T^{(ab)}_{ij}\simeq-2\mu^2\delta_{ij}\psi_a\psi_b$, which gives $h^{(ab)}_{00}\simeq12\pi G\,\psi_a\psi_b$. This is all we need, as only the $h^{00}\partial_t^2$ term of $\delta\Box^{(ab)}$ contributes at this order. The result is an effective pointlike interaction, which shifts the SI coupling by
\begin{equation}
    \lambda_{Gs} = 12\pi G \mu^2\,.
\end{equation}

To simplify at PN order, we use $\dot{\mathbf a}^{(bc)}_{\rm BH}=-i\omega_{bc}\,\mathbf a^{(bc)}_{\rm BH}$, and set $\omega\to\mu$ in terms that are already $\mathcal O(v^2)$. We split the scattering source of Eq.~\eqref{eq:S} into
\begin{equation}
\begin{aligned}
    S_{abc} &= \zeta_{abc}S^{\rm N}_{abc}+S^{h_{ij}}_{abc}+S^{\rm ret}_{abc}+S^{\rm GM}_{abc} \\
    &\quad +S^{\rm frame}_{abc}+S^{G^2}_{abc}+(a\leftrightarrow b)+S^{s}_{abc}\,,
\end{aligned}
\label{eq:SPN}
\end{equation}
where
\begin{equation}
    S^{\rm N}_{abc} = \mu^2\big(\Phi^{(bc)}_{\rm N}+\mathbf a^{(bc)}_{\rm N}\!\cdot\bx\big)\psi_a
\end{equation}
is the Newtonian source with recoil,
\begin{equation}
\begin{aligned}
    \zeta_{abc} &\equiv \frac{\omega_a^2\left(2\omega_b\omega_c-\mu^2\right)}{\mu^4} \\
    &\simeq 1-\alpha^2\left(\frac{1}{n_a^2}+\frac{1}{n_b^2}+\frac{1}{n_c^2}\right)\,,
\end{aligned}
\end{equation}
and
\begin{align}
    S^{h_{ij}}_{abc} &= -\Phi^{(bc)}_{\rm N}\nabla^2\psi_a\,, \\
    S^{\rm ret}_{abc} &= \mu^2\big(\Phi^{(bc)}_{\rm ret}+\mathbf a^{(bc)}_{\rm ret}\!\cdot\bx\big)\psi_a\,, \\
    S^{\rm GM}_{abc} &= -4i\mu\,\mathbf A^{(bc)}\!\cdot\nabla\psi_a + \mu^2\big(\mathbf a^{(bc)}_{\rm GM}\!\cdot\bx\big)\psi_a\,, \\
    S^{\rm frame}_{abc} &= \tfrac12\mu\,\omega_{bc}\big(\mathbf a^{(bc)}_{\rm N}\!\cdot\bx\big)\psi_a - \tfrac12\,\mathbf a^{(bc)}_{\rm N}\!\cdot\nabla\psi_a\,, \\
    \begin{split}
        S^{G^2}_{abc} &= \mu^2\Big(\Phi^{(bc)}_G+\mathbf a^{(bc)}_G\!\cdot\bx+\Phi_{\rm BH}\Phi^{(bc)}_{\rm N}(0) \\
        &\qquad -2\Phi_{\rm BH}\Phi^{(bc)}_{\rm N}-\tfrac52\Phi_{\rm BH}\,\mathbf a^{(bc)}_{\rm N}\!\cdot\bx\Big) \psi_a\,,
    \end{split} \\
    S^{s}_{abc} &= -\tfrac12\lambda_{Gs}\,\psi_a\psi_b\bar\psi_c\,.
\end{align}
The source is inserted into Eq.~\eqref{eq:master} exactly as in the relativistic case. The PN corrections are:
\begin{itemize}
    \item $\zeta_{abc}$: binding energy gravitates, and couples to gravity. The pair $(bc)$ sources $\Phi$ with its energy rather than its rest mass, and $\Phi$ acts on the energy of $\ket{a}$, so the binding energies of all three levels rescale the Newtonian amplitude.
    \item $S^{h_{ij}}$: the spatial curvature $h_{ij}$ couples to the gradient of the field.
    \item $S^{\rm ret}$: the finite travel time of the graviton across the cloud.
    \item $S^{\rm GM}$: the mass current sources $\mathbf A$, which drags the field and accelerates the BH.
    \item $S^{\rm frame}$: corrections to the fictitious force in the accelerated BH frame. Together with the terms in $\mathbf{a}_{\rm BH}$, these are effectively part of the recoil.
    \item $S^{G^2}$: the exchange takes place in the potential of the BH, which enters through the background metric in the density, the nonlinearity of gravity, the redshift of the BH's mass, and the retardation of the BH's own potential.
    \item $S^{s}$: $s$-channel exchange.
\end{itemize}

\begin{figure*}
    \centering
    \includegraphics[width=0.98\linewidth]{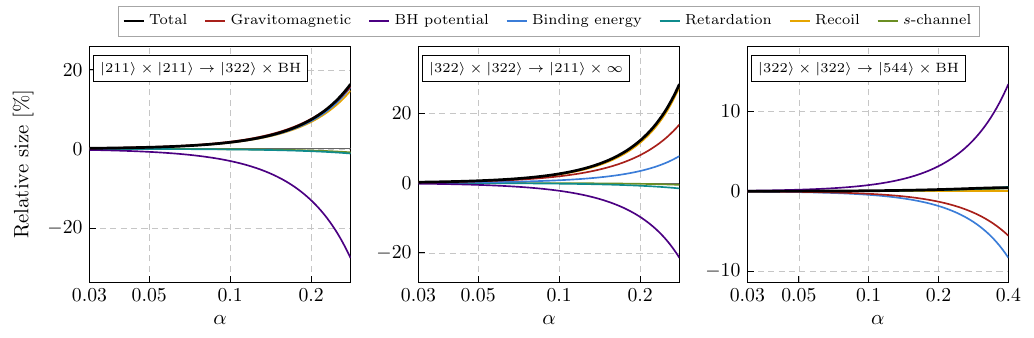}
    \caption{Relative size, in percent, of the leading PN corrections to the SG amplitude of three scattering channels for $\tilde a=0.95$.}
    \label{fig:pn_corr}
\end{figure*}

In Fig.~\ref{fig:pn_corr} we show a comparison of the different PN contributions to the scattering processes. The error from these corrections, which we do not include in our evolution runs, is up to $\mathcal{O}(10\%)$ for the values of $\alpha$ where these states are long lived. This remains true even as $\alpha$ increases, because only high-$n$ states become long-lived. These states have a small virial velocity, $v^2 \propto n^{-2}$ (as the peak of their wavefunctions sits far from the BH horizon), so the error in the matrix elements stays moderate and does not affect our conclusions. As emphasized in the main text, a large relativistic correction comes from the decay rates of the spherical states, which we include throughout our calculations (see Fig.~\ref{fig:gf_vs_hydrogenic}).  We include the PN corrections here mainly to show that the Newtonian treatment of the self-gravity matrix elements is well behaved at appropriate values of $\alpha$. The complexity of even the leading PN contributions suggests that any precision calculation is better served by the fully relativistic approach outlined in Sec.~\ref{sec:relativistic}, where the PN expansion can serve for the convergence check.

\subsection{Black-hole recoil}
\label{sec:recoil}
As emphasized in the previous section the BH is not at rest, as the cloud and BH move around their common center of mass, so the BH carries a momentum opposite and equal to that of the cloud. In order to set up the fully relativistic calculation we moved to the (non-interital) BH-centered frame, while in the main text we work in the (intertial) atom-centered frame. This allows for an action formulation and a more straightforward inclusion of other interactions, \eg the tidal coupling. Let us now derive the recoil kernel of Eq.~\eqref{eq:Vrec}.

In the atom-centered frame, the Lagrangian that describes both the cloud and the BH  degrees of freedom is
\begin{equation}
L=\tfrac12M\dot{\mathbf X}^2+\int_{\tilde\bx}\bar\psi\Big(i\partial_t+\frac{\nabla^2}{2\mu}
 +\frac{\alpha}{|\tilde\bx-\mathbf X|}\Big)\psi+L_{\rm quart}[\psi]\,,
\label{eq:Lrec}
\end{equation}
where $\mathbf X$ is the position of the BH, $\tilde\bx$ is the coordinate of the field, and $L_{\rm quart}$ contains the SI and the Newtonian SG of the cloud. We now measure the field from the BH, $\bx\equiv\tilde\bx-\mathbf X$. At fixed $\bx$ the time derivative becomes $\partial_t\psi\to\partial_t\psi-\dot{\mathbf X}\cdot\nabla\psi$, so the first two terms of Eq.~\eqref{eq:Lrec} turn into the Lagrangian of the atom with the BH at the origin, plus
\begin{equation}
L_{\rm rec}=\tfrac12M\dot{\mathbf X}^2+\dot{\mathbf X}\cdot\mathbf P_\psi\,.
\end{equation}
The position $\mathbf X$ no longer appears, only $\dot{\mathbf X}$, so its conjugate momentum $\mathbf P_{\mathbf X}=M\dot{\mathbf X}+\mathbf P_\psi$, the total momentum, is conserved. In the center-of-mass frame it vanishes, and $\dot{\mathbf X}=-\mathbf P_\psi/M$. We eliminate the BH with the Routhian $R=L-\mathbf P_{\mathbf X}\cdot\dot{\mathbf X}$, which acts as a Lagrangian for the cloud alone. This gives $R_{\rm rec}=-\mathbf P_\psi^2/(2M)$, hence $H_{\rm rec}=\mathbf P_\psi^2/(2M)$, which is Eq.~\eqref{eq:Vrec}.

The matrix element~\eqref{eq:U} can be further expressed as 
\begin{eqnarray}
U^{\rm rec}_{abcd}&=&\frac{\langle a|\hat{\mathbf p}|b\rangle\cdot\langle c|\hat{\mathbf p}|d\rangle}{M} \, \label{eq:Urec} \\
&=&-\frac{\mu^2}{M}\,(E_a-E_b)(E_c-E_d)\,\langle a|\bx|b\rangle\cdot\langle c|\bx|d\rangle\,, \nonumber
\end{eqnarray}
where the second line uses $\langle a|\hat{\mathbf p}|b\rangle=i\mu\,(E_a-E_b)\langle a|\bx|b\rangle$. Two properties used in the main text follow from Eq.~\eqref{eq:Urec}. First, recoil is purely dipolar. Each pair of levels must satisfy $\Delta\ell=\pm1$ and $\Delta m=0,\pm1$, with $\Delta m_{ab}=-\Delta m_{cd}$. It therefore contributes to, \eg channels such as  $\ket{211}\times\ket{211}\to\ket{322}\times{\rm BH}$ and $\ket{322}\times\ket{322}\to\ket{211}\times\infty$, but not to, \eg $\ket{322}\times\ket{322}\to\ket{544}\times{\rm BH}$. Second, its size is set by the energy differences. For levels of different $n$ these are of order $\mu\alpha^2$, and recoil is as large as the Newtonian term, $N_aU^{\rm rec}\sim\mu\alpha^2M_{\rm c}/M$. For levels of the same $n$, the differences are the fine and hyperfine splittings, and the recoil is suppressed by $\alpha^2$ and $\tilde a\alpha^3$ per pair.
\subsection{Two-level dynamics} \label{app:two_state}
Here we derive the three evolutionary stages for two-level systems outlined in the main text. We consider levels $\ket{a}$ and $\ket{b}$ with $m_b=2m_a$, and write $x\equiv\varepsilon_a$ and $y\equiv\varepsilon_b$ for their occupations. The relevant processes,\footnote{Their content  can be understood also from the energy conservation perspective. Focusing first on $\{\ket{211},\ket{322}\}$, in $\ket{211} \times \ket{211} \to \ket{322} \times \ket{\psi_4}$,  the fourth particle goes to a bound  state as $\mu(1-7\alpha^2/36) < \mu$, and as $m_{\psi_4}=0$ it falls quickly into the BH. In contrast for $\ket{322} \times \ket{322} \to \ket{211} \times \ket{\psi_4}$ we have $\mu(1+\alpha^2/72) > \mu$, and the particle escapes to infinity (see also~\cite{Baryakhtar:2020gao}).} explained in the main text, lead to the following evolution equations:
\begin{equation}
\begin{aligned}
\dot x&=\gamma^{\rm SR}_a\,x-2\gamma_{\rm BH}\,x^2y+\gamma_\infty\,xy^2+\gamma_{\rm tr}\,xy\,,\\
\dot y&=\gamma^{\rm SR}_b\,y+\gamma_{\rm BH}\,x^2y-2\gamma_\infty\,xy^2-\gamma_{\rm tr}\,xy\,,
\end{aligned}
\label{eq:kin}
\end{equation}
In the following we will consider first a general case, while neglecting GW annihilation that acts only on long timescales (see the end of this section).

\emph{(i) Saturation.} Mixing processes in Eq.~\eqref{eq:kin} can stop the growth of $\ket{a}$ by reaching a fixed point, $\dot{x}=\dot{y}=0$. Assuming $\gamma^{\rm SR}_b \ll \gamma^{\rm SR}_a$, this gives
\begin{eqnarray}
x_\mathrm{fp}&=&\tfrac13\left[2\varepsilon_\star+\sqrt{\varepsilon_\star^2+9\,\varepsilon_{\rm eq}^2}\,\right],\qquad y_\mathrm{fp}=\frac{\gamma_{\rm BH}x-\gamma_{\rm tr}}{2\gamma_\infty}\,,  \nonumber \\
\varepsilon_{\rm eq}&\equiv& \left(\frac{4\gamma_\infty\gamma^{\rm SR}_a}{3\gamma_{\rm BH}^2}\right)^{1/2}\,, \label{eq:plateau}
\end{eqnarray}
with $\varepsilon_\star=\gamma_{\rm tr}/\gamma_{\rm BH}$ of Eq.~\eqref{eq:relic}. The fixed point lies above both $\varepsilon_\star$ and $\varepsilon_{\rm eq}$. Scattering can therefore cap $\ket{a}$ only if this $x_\mathrm{fp} < \varepsilon_a^{\rm sat}$. This condition is necessary but not sufficient: superradiance can drive $x$ past the fixed point, up to saturation, while $y$ is still small. The $y$-dependent terms in the first line of Eq.~\eqref{eq:kin} must also become comparable to the superradiant term, which requires
\begin{equation}
y>\frac{\gamma^{\rm SR}_a}{2\gamma_{\rm BH}\,x-\gamma_{\rm tr}}\,.
\label{eq:ycap}
\end{equation}
If either condition fails, $x$ grows to its free-field value.

The later evolution depends on whether the fixed point is reached. The SI-dominated regime has been studied extensively~\cite{Baryakhtar:2020gao,Witte:2024drg,Cannizzaro:2026qgf}, so we focus on $f_a \gtrsim f_a^\star$. There, for all $\{\alpha_0,\tilde{a}_0\}$ for which $\ket{211}$ is superradiant, we find that Eq.~\eqref{eq:ycap} is not satisfied, and $\ket{211}$ saturates as for a free field. For higher-$n$ states the fixed point is reached. We consider the two cases in turn.

\emph{(ii) Depletion of $\{\ket{211},\ket{322}\}$.} After saturation the BH sits at the superradiant threshold of $\ket{a}$, so $\gamma^{\rm SR}_a\simeq0$. For now we also neglect $\gamma^{\rm SR}_b$ and $\gamma_{\rm tr}$, as both matter only at the end of the depletion, and we restore them in \emph{(iii)}.

What is left of Eq.~\eqref{eq:kin} depends on the two rates only through their ratio, $\beta\equiv\gamma_{\rm BH}/\gamma_\infty$. The ratio of the occupations, $r\equiv y/x$, obeys
\begin{equation}
\dot r=\gamma_\infty\,x^2\,r\left[\beta+2(\beta-1)\,r-r^2\right].
\label{eq:rdot}
\end{equation}
The bracket is a downward parabola in $r$ that is positive at $r=0$, so it has exactly one positive root,
\begin{equation}
r_\star=\beta-1+\sqrt{\beta^2-\beta+1}\simeq
\begin{cases}
    \tfrac12\beta & \text{for $\beta \ll 1$\,,} \\
    2\beta & \text{for $\beta \gg 1$\,,}
\end{cases}
\label{eq:rstar}
\end{equation}
which is an attractor. For $r<r_\star$ the bracket is positive, and $\ket{a}\times\ket{a}\to\ket{b}\times{\rm BH}$ moves particles from $\ket{a}$ to $\ket{b}$. For $r>r_\star$ it is negative, and $\ket{b}\times\ket{b}\to\ket{a}\times\infty$ moves them back. Taking $r\simeq r_\star$ in the first line of Eq.~\eqref{eq:kin}, we obtain
\begin{align}
x&=\left[(\varepsilon_a^{\rm sat})^{-2}+2C(t-t_{\rm sat})\right]^{-1/2}, \label{eq:Cexact} \\
C&=\gamma_\infty\,r_\star(2\beta-r_\star)\simeq
\begin{cases}
    3\gamma_{\rm BH}^2/(4\gamma_\infty) & \text{for $\beta \ll 1$\,,} \\
    3\gamma_{\rm BH} & \text{for $\beta \gg 1$\,,}
\end{cases} \nonumber
\end{align}
where $t_{\rm sat}$ is the time of saturation\footnote{The $\ell=0$ particles absorbed by the BH carry mass but no angular momentum, so they lower $\tilde a$ and push the BH below the threshold of $\ket{211}$. If scattering is fast, $\ket{211}$ is then reabsorbed; our runs suggest this occurs for $\alpha_0\gtrsim0.15$. For lower $\alpha$ scattering is slower, and this spin-down is offset by the absorption of $\ket{211}$, which spins the BH back up and keeps it at threshold. In effect this slightly increases the depletion of $\ket{211}$, $C\to C(1+8\alpha^2)$.}. At late times this is Eq.~\eqref{eq:depletion} (here, however, we do not assume $\beta$ is small).

\begin{figure}
    \centering
    \includegraphics[width=0.99\linewidth]{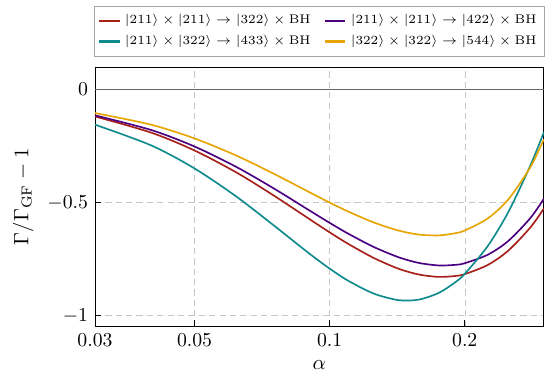}
    \caption{Scattering rates from the hydrogenic expansion (flat-space $\ket{n00}$ levels and continuum) relative to those computed with the Kerr Green's function, $\Gamma_{\rm GF}$, for four channels whose fourth particle is absorbed by the BH, at $\tilde a=0.95$.}
    \label{fig:gf_vs_hydrogenic}
\end{figure}

We first take the SG-dominated regime, $f_a \gg f_a^\star$. A simple estimate gives the $\alpha$ scaling of $C$. Two particles in the cloud interact with an energy $U_\mathrm{sg}\sim G\mu^2/r_\mathrm{B}\sim G\mu^3\alpha$. Levels are spaced by $\Delta\sim\mu\alpha^2$, and an $\ell=0$ state is absorbed at a rate $\Gamma_{\ell=0}\sim\mu\alpha^5$. The rate per pair $\ket{a}\times\ket{a}$ is then $\gamma_\mathrm{BH}\sim(GM^2)^2\sum\Gamma_{\ell=0}\,(U_\mathrm{sg}/\Delta)^2\sim\mu\alpha^7$, where the factor $(GM^2)^2$ converts occupation numbers to $\varepsilon=N/(GM^2)$. For $\gamma_\infty$, the escaping particles have momenta $k\sim\mu\alpha$, while the cloud is spread over $\Delta k\sim r_\mathrm{B}^{-1}$ in momentum space. Fermi's golden rule then gives the rate per pair $\ket{b}\times\ket{b}$, $\gamma_\infty\sim(GM^2)^2(\mu/k)\,U_\mathrm{sg}^2/\Delta k\sim\mu\alpha^4$. Consequently, $C=c_{\rm SG}\,\alpha^{11}/(GM)$, where $c_{\rm SG}$ depends on the dimensionless kernels. At leading order in $\alpha$, the two rates are
\begin{equation}
\gamma_{\rm BH}^{\rm LO}=0.067\,\tilde r_+\,\alpha^7\mu\,,\qquad
\gamma_{\infty}^{\rm LO}=5.6\times10^{-3}\,\alpha^4\mu\,,
\label{eq:ratesLO}
\end{equation}
with $\tilde r_+\equiv1+\sqrt{1-\tilde a^2}$, which gives $c_{\rm SG}=0.60\,\tilde r_+^2\simeq2.3$. This underestimates the depletion. As explained in the main text, our evolutions use rates computed with Kerr quasi-bound states and the Kerr Green's function, which changes the $\ell=0$ intermediate states at $\mathcal{O}(1)$ (Fig.~\ref{fig:gf_vs_hydrogenic}). Table~\ref{tab:cSG} lists the resulting $c_{\rm SG}$, which is the value that enters our evolutions and Eq.~\eqref{eq:Mc}; see the caption for its dependence on spin. It is larger than $2.3$ for every $\alpha$ in the table, mainly because $\gamma_{\rm BH}$ is up to five times its leading-order value.

\begin{table}
\caption{The rate $\gamma_{\rm BH}$ relative to its leading-order value, the occupation ratio $r_\star$ of Eq.~\eqref{eq:rstar}, and the depletion coefficient $c_{\rm SG}\equiv C\,GM/\alpha^{11}$ of Eq.~\eqref{eq:Cexact} for SG alone. All three follow from the scattering rates used in our evolutions, which are computed at $\tilde a=0.95$ and rescaled with $\tilde r_+$ to the spin at which $\ket{211}$ saturates. Evaluating $\gamma_{\rm BH}$ directly at that spin lowers it by a factor of $1.1$--$1.5$, and $c_{\rm SG}$ by a factor of $1.2$--$2.4$.}
\label{tab:cSG}
\begin{ruledtabular}
\begin{tabular}{cccc}
$\alpha$ & $\gamma_{\rm BH}/\gamma_{\rm BH}^{\rm LO}$ & $r_\star$ & $c_{\rm SG}$\\
\hline
0.03 & 1.2 & $4.3\times10^{-4}$ & 3.8\\
0.04 & 1.3 & $1.1\times10^{-3}$ & 4.3\\
0.05 & 1.4 & $2.3\times10^{-3}$ & 5.2\\
0.06 & 1.6 & $4.6\times10^{-3}$ & 6.5\\
0.075 & 1.9 & $1.1\times10^{-2}$ & 9.5\\
0.09 & 2.3 & $2.4\times10^{-2}$ & 15\\
0.10 & 2.6 & $3.9\times10^{-2}$ & 20\\
0.12 & 3.5 & $9.6\times10^{-2}$ & 35\\
0.15 & 4.8 & 0.31 & 71\\
0.175 & 5.3 & 0.63 & 84\\
0.20 & 4.7 & 0.88 & 63\\
0.225 & 3.7 & 0.97 & 36\\
0.25 & 2.8 & 1.0 & 19\\
\end{tabular}
\end{ruledtabular}
\end{table}

We now include SI. Then $\gamma_{\rm BH}$ and $\gamma_\infty$ also depend on $f_a$, through Eq.~\eqref{eq:gamma}. Both channels that involve $\ket{211}$ have $\kappa=-1$, so SI and SG interfere destructively around $f_a^\star$, which differs between the two channels. Close to $f_a^\star$ of $\ket{211}\times\ket{211}\to\ket{322}\times{\rm BH}$, $\gamma_{\rm BH}$ nearly vanishes. Since $C\propto\gamma_{\rm BH}^2$ and $x\propto C^{-1/2}$, the cloud depletes more slowly and is heavier at a given age, by up to a factor of two in our runs (Fig.~\ref{fig:evolution_si}). Close to $f_a^\star$ of $\ket{322}\times\ket{322}\to\ket{211}\times\infty$, SI speed up the depletion. There $\beta$ is no longer small and the exact expressions are needed, as in the middle panel of Fig.~\ref{fig:evolution}. For even lower $f_a$ the depletion is faster still, and one enters the pure SI regime~\cite{Baryakhtar:2020gao,Witte:2024drg,Cannizzaro:2026qgf}.

Finally, the rates are sums of squares of the combined SG and SI amplitudes, $\gamma\propto\sum_\ell\big[A^{\rm SG}_\ell+(M_{\rm Pl}/f_a)^2A^{\rm SI}_\ell\big]^2$. The sum runs over the angular momenta of the fourth particle, which is a quasi-bound state for $\gamma_{\rm BH}$ and a continuum state for $\gamma_\infty$, and $A_\ell$ are the corresponding amplitudes. In general, therefore, $\gamma(f_a)=\gamma_{\rm SG}+2\kappa\sqrt{\gamma_{\rm SG}\gamma_{\rm SI}}\,(M_{\rm Pl}/f_a)^2+\gamma_{\rm SI}\,(M_{\rm Pl}/f_a)^4$, where $|\kappa|\leq1$ is the normalized overlap of the SG and SI amplitudes, and Eq.~\eqref{eq:gamma} holds only for $\kappa=\pm1$. For the BH channels, $\ell=0$ dominates and higher multipoles are suppressed by powers of $\alpha$, so $\kappa=\pm1$ to leading order in $\alpha$. For the channels to infinity several multipoles can contribute. For example, $1-|\kappa|\simeq10^{-3}$ for $\ket{322}\times\ket{322}\to\ket{211}\times\infty$, where the $\ell=3$ and $\ell=5$ continuum states contribute. Our numerical evolutions use the full expression.

\emph{(iii) Remnant of $\{\ket{211},\ket{322} \}$.} The depletion law cannot hold forever, because the terms we neglected eventually matter: the GW transition, the superradiance of $\ket{b}$, and GW annihilation. We now estimate when. With the first two, the second line of Eq.~\eqref{eq:kin} reads $\dot y=x\,y\,(\gamma_{\rm BH}x-\gamma_{\rm tr}-2\gamma_\infty y+\gamma^{\rm SR}_b/x)$. GW annihilation of $\ket{a}$ adds $-\gamma_{\rm ann}x^2$ to the first line. The depletion law therefore breaks down once $x$ has dropped to
\begin{equation} \label{eq:xend}
x_{\rm end}=\max\left[ \varepsilon_\star \,,\left(\frac{\gamma^{\rm SR}_b}{\gamma_{\rm BH}}\right)^{1/2} , \frac{4\gamma_\infty\gamma_{\rm ann}}{3\gamma_{\rm BH}^2}\right]\,.
\end{equation}
With Eq.~\eqref{eq:Cexact} at late times, $t\simeq1/(2Cx^2)$, this happens at
\begin{equation} \label{eq:tend}
t_{\rm end}=\min\left[\frac{\gamma_{\rm BH}^2}{2C\gamma_{\rm tr}^2}\,,\frac{\gamma_{\rm BH}}{2C\gamma^{\rm SR}_b}\,,\frac{9\gamma_{\rm BH}^4}{32C\gamma_\infty^2\gamma_{\rm ann}^2}\right].
\end{equation}

For $\{\ket{211},\ket{322}\}$ the second entry always wins. The GW transition $\ket{322}\to\ket{211}$ has $\Delta\ell=1$, which parity forbids for the mass quadrupole. It proceeds only through the mass octupole and the current quadrupole, with $\gamma_{\rm tr}\simeq 5 \times10^{-6}\, \mu\, \alpha^{10}$~\cite{Baryakhtar:2020gao}. Since $\gamma^{\rm SR}_{322}\propto \mu\, \alpha^{13}$, the first two entries of Eq.~\eqref{eq:xend} both scale as $\alpha^3$, but $\gamma_{\rm tr}/\gamma_{\rm BH}$ lies about three orders of magnitude below $(\gamma^{\rm SR}_{322}/\gamma_{\rm BH})^{1/2}$. For annihilation, $\gamma_{\rm ann}\propto  \mu\, \alpha^{14}$ for $\ket{211}$~\cite{Brito:2017zvb,Yoshino:2013ofa}, so the third entry scales as $\alpha^4$ and is suppressed by one more power of $\alpha$. The superradiant growth of $\ket{322}$ therefore ends the depletion stage of Eq.~\eqref{eq:depletion}, at $t_{\rm end}\simeq2\gamma_\infty/(3\gamma_{\rm BH}\gamma^{\rm SR}_{322})$.

\emph{Beyond the two-state approximation.} Eq.~\eqref{eq:kin} assumes that no third level takes part. While $\ket{211}$ grows and saturates, $\ket{211}\times\ket{211}\to\ket{n22}\times{\rm BH}$ feeds all overtones $\ket{n22}$, but $\ket{322}$ has the largest rate and becomes macroscopic first, while the others stay far below it. Once the $\{\ket{211},\ket{322}\}$ pair has formed, a third level $\ket{c}$ is fed by the pair but also drained by it, through cross channels such as $\ket{n22}\times\ket{322}\to\ket{211}\times\infty$, and by its GW transitions. As long as its occupation $z\equiv\varepsilon_c$ is small, it obeys a linear equation, $\dot z \approx \gamma_{\rm eff}(x,y)\,z$. Integrating $\gamma_{\rm eff}$ over the depletion stage, Eq.~\eqref{eq:Cexact}, we find that none of $\ket{n22}$, $\ket{n33}$ and $\ket{n44}$ grows appreciably, so $\{\ket{211},\ket{322}\}$ stays closed until $t_{\rm end}$. 

This is no longer the case at later stages. After $\ket{211}$ is reabsorbed, the growth of $\ket{322}$ stalls at the fixed point of Eq.~\eqref{eq:plateau}, with $\ket{544}$ as partner, and once the BH reaches the threshold of $\ket{322}$, $x_{\rm fp}$ reduces to $\varepsilon_\star$ of Eq.~\eqref{eq:relic}. This pair, however, need not be stable: higher overtones $\ket{n44}$ can grow on top of it and take over as the partner, in the burst-like cadence (cf. Fig.~\ref{fig:evolution}). The system is then still approximately a two-state one, $\{\ket{322},\ket{n44}\}$, with a partner that changes in time. Once the evolution proceeds to higher pairs, such as $\{\ket{433},\ket{n66}\}$, and levels with $m=5$, $6$ and $8$ are populated together, not even an approximate two-state description holds, and no simple semi-analytic description is available.
\subsection{Varying the parameters}
Figure~\ref{fig:evolution} of the main text shows three cases. Here, we provide additional examples. All runs evolve the 215 levels from $\tilde a_0=0.95$ for $10^9$\,yr, with relativistic GW rates. For clarity, we only show levels that carry more than $2\times10^{-8}M_0$ at some point in the evolution.
\subsubsection{Effect of self-gravity}
\begin{figure*}
    \centering
    \includegraphics[width=0.99\linewidth]{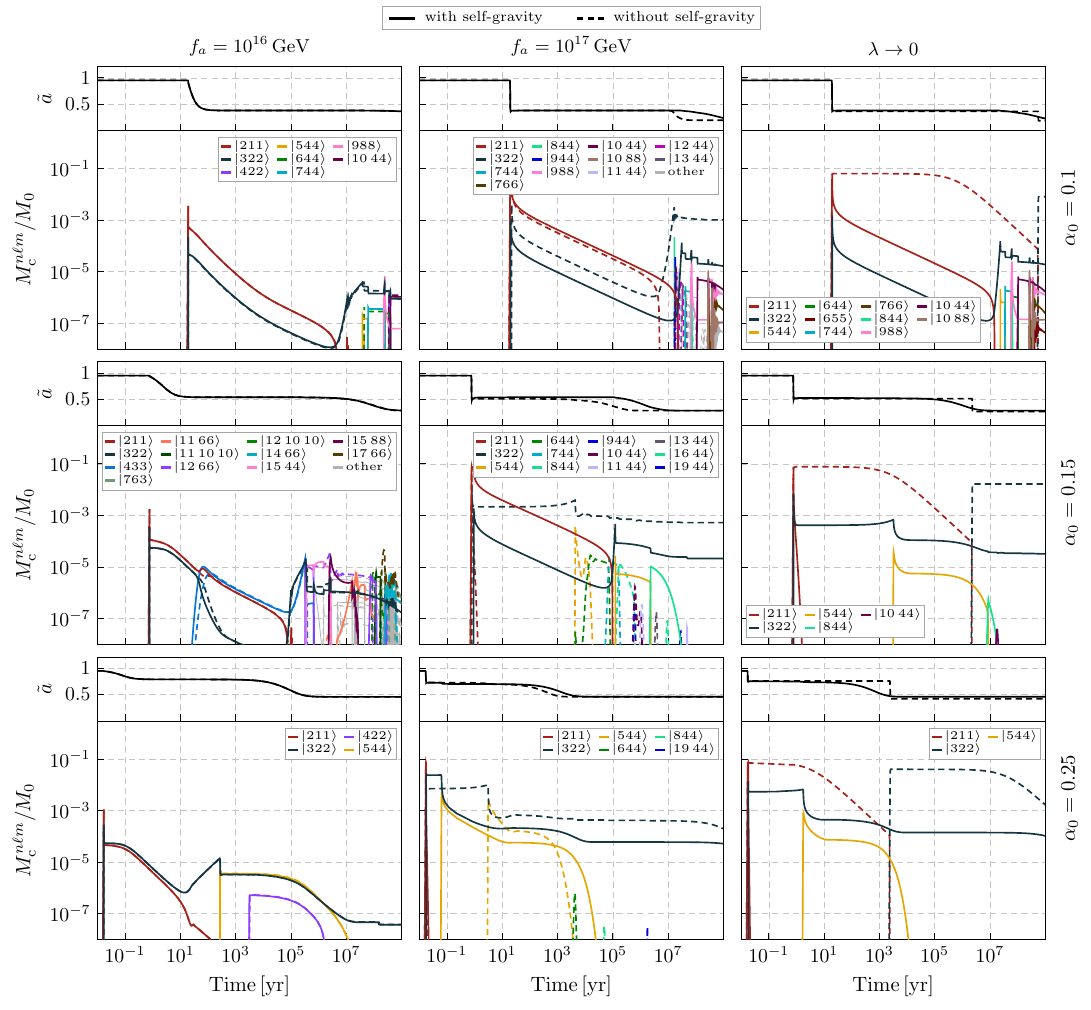}
    \caption{Evolution with (solid) and without (dashed) SG for $M_0=10\,M_\odot$: BH spin (top) and cloud mass per level (bottom). SIs are included in both sets of curves; for $\lambda\to0$ (right column), \ie without SI, the dashed curves are the free field.}
    \label{fig:evolution_sm10}
\end{figure*}

Figure~\ref{fig:evolution_sm10} compares runs that differ only in whether SG is included, for a $10\,M_\odot$ BH.

Without SI (right column) the dashed curves are the free field, and SG removes most of the cloud. For $\alpha_0=0.1$ the cloud depletes as $t^{-1/2}$, and at intermediate times is orders of magnitude lighter than without SG. For $\alpha_0\geq0.15$, $\ket{211}$ is reabsorbed right after saturation and $\ket{322}$ inherits the cloud and falls to its remnant once $\ket{544}$ has grown.

The cluster of high levels that appears after $10^7$\,yr in the top right panel is the cascade described in the main text. Such towers have been found before in the context of SI, \cite{Cannizzaro:2026qgf} studies their convergence. Here they appear even without SI.

For $f_a=10^{17}$\,GeV (middle column) SI alone already deplete the cloud, so the two sets of curves are closer, but they do not coincide. At $\alpha_0=0.1$ the two interactions interfere and the cloud is actually heavier with SG included (while at late times, the order is reversed again).

For $f_a=10^{16}$\,GeV (left column) SI dominate every channel and SG changes little. The final spins agree to within $10^{-3}$. The cloud masses agree to within $10\%$ for $\alpha_0=0.1$ and $0.25$, and to within a factor of two for $\alpha_0=0.15$, where many high levels are populated in both runs.
\subsubsection{Dependence on the black-hole mass}
\begin{figure*}
    \centering
    \includegraphics[width=0.99\linewidth]{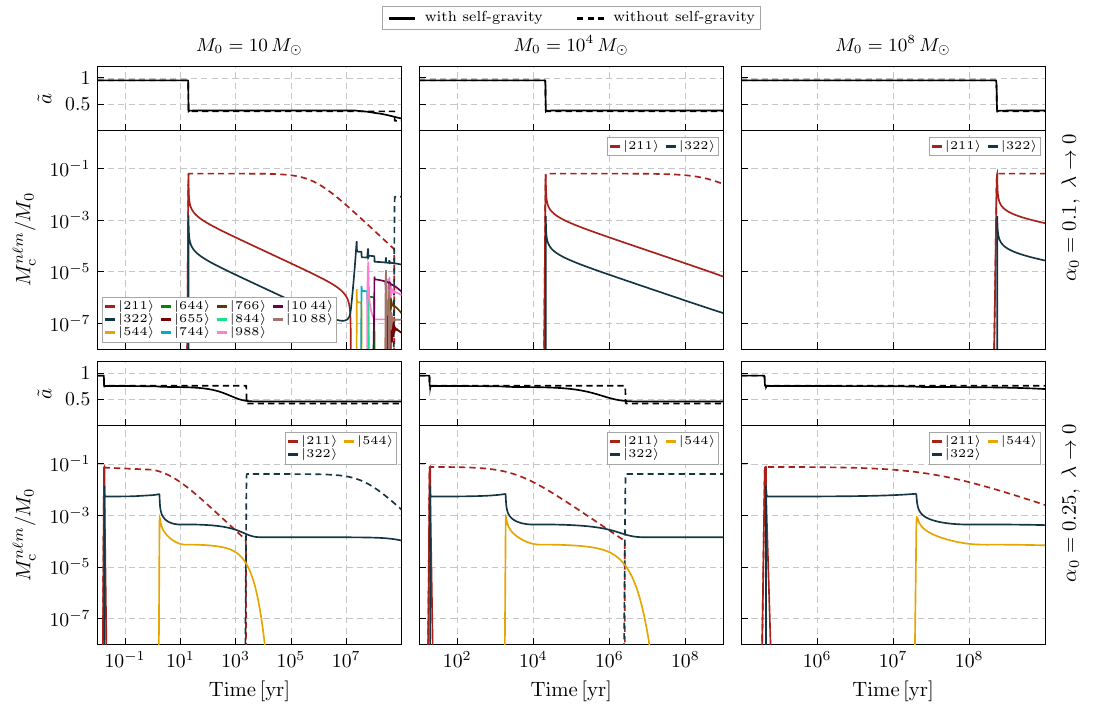}
    \caption{Evolution with (solid) and without (dashed) SG for three BH masses, without SI ($\lambda\to0$); the dashed curves are the free field.}
    \label{fig:evolution_mass}
\end{figure*}

At fixed $\alpha_0$ and $f_a$ every rate scales as $1/M_0$, so heavier BHs follow the same evolution with time rescaled (Fig.~\ref{fig:evolution_mass}). A run of $10^9$\,yr therefore covers less of the evolution for a heavier BH. For $\alpha_0=0.1$ the $10\,M_\odot$ run shows the depletion of $\ket{211}$, its reabsorption and the bursts that follow; the $10^4\,M_\odot$ run shows only the $t^{-1/2}$ depletion, and the $10^8\,M_\odot$ run only its beginning. For $\alpha_0=0.25$ all three masses go through the same sequence: $\ket{211}$ is reabsorbed, $\ket{322}$ takes over, $\ket{544}$ grows, and $\ket{322}$ settles at a remnant of order $10^{-4}M_0$. For $10^8\,M_\odot$ no level beyond $\ket{544}$ is populated within $10^9$\,yr. The runs with SI scale in the same way.
\subsubsection{Effect of self-interactions}
\begin{figure*}
    \centering
    \includegraphics[width=0.99\linewidth]{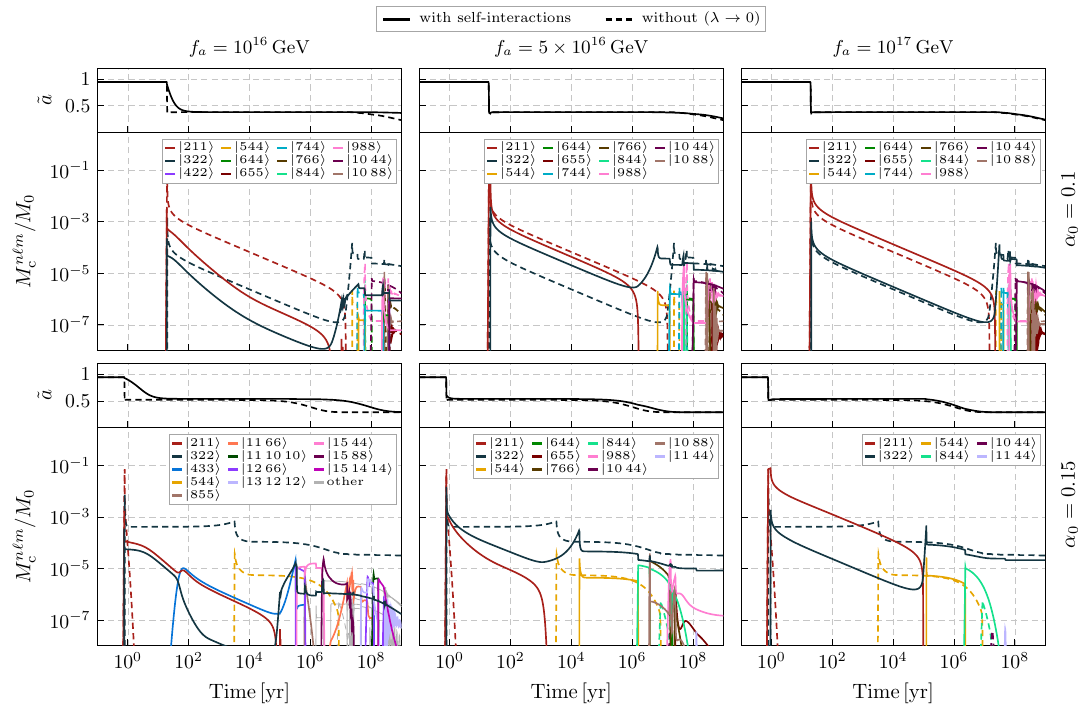}
    \caption{Evolution with SG, with (solid) and without (dashed) SI, for $M_0=10\,M_\odot$. The dashed curves are the same across each row.}
    \label{fig:evolution_si}
\end{figure*}

Figure~\ref{fig:evolution_si} shows the converse comparison: SG is always included (as it should be), and the decay constant is varied, for $\alpha_0=0.1$ and $0.15$.

In the top row ($\alpha_0=0.1$) the cloud with SI goes from lighter to heavier than the one with SG alone as $f_a$ increases. On the left, SI dominate~\eqref{eq:fstar}. They stop the growth of $\ket{211}$ well before saturation, as is known~\cite{Baryakhtar:2020gao,Witte:2024drg,Cannizzaro:2026qgf}, and then deplete it faster, so the solid curves lie far below the dashed ones. In the middle the two interactions are comparable: the cloud saturates as with SG alone, but still depletes somewhat faster and $\ket{211}$ is reabsorbed earlier. On the right the order is reversed. The solid curves now lie above the dashed ones, the cloud is twice as heavy, and $\ket{211}$ lives longer. This reversal is the interference of Eq.~\eqref{eq:gamma}: below the decay constant at which $\ket{211}\times\ket{211}\to\ket{322}\times{\rm BH}$ cancels, SI add to the depletion, and above it they partly cancel SG. The bursts of high levels at late times occur in all three panels.

The bottom row ($\alpha_0=0.15$) shows the same trend more sharply. With SG alone, $\ket{211}$ is reabsorbed right after saturation, so its dashed curve is only a spike and $\ket{322}$ carries the cloud. On the left, SI again cap the cloud and spread it over many high levels. In the middle, $\ket{211}$ lasts a little longer than the spike. On the right it survives for $2\times10^4$\,yr and the cloud is heavier than with SG alone, because this decay constant lies close to the one at which the channel cancels. With SI alone at the same $f_a$, $\ket{211}$ is reabsorbed at once as well (middle panel of Fig.~\ref{fig:evolution_sm10}).

This cancellation deserves a final remark. For a channel with $\kappa=-1$ the SG and SI amplitudes have opposite signs, and the rate of Eq.~\eqref{eq:gamma} vanishes at $f_a=f_a^\star$. This does not bring back the free field, for two reasons. First, the zero is narrow: the rate is below a tenth of its SG value only for $f_a$ between about $0.9$ and $1.2$ $f_a^\star$ and, since $f_a^\star\propto\alpha$, only in an equally narrow range of $\alpha$ at fixed $f_a$. Second, $f_a^\star$ differs from channel to channel (Fig.~\ref{fig:cross-over}), so no decay constant switches off all of them: where one channel cancels, the others still scatter. The cancellation therefore shows up as a cloud that is heavier and a $\ket{211}$ that lives longer, as in the right column of Fig.~\ref{fig:evolution_si}, not as a return to the free field. Neither interaction can be dropped in favor of the other.
\subsection{Clouds in binaries}
Here we give the Hamiltonian used for the binary of the main text and explain how the companion acts on the cloud. The orbit adds its mean anomaly $\vartheta$ and frequency $\Omega$ to the phase space, and the Hamiltonian gains a Keplerian term and a tidal coupling~\cite{Boskovic:2025ixx},
\begin{equation}
\begin{aligned}
H&\supset-\frac{qM}{2(1+q)}\bigl[GM(1+q)\,\Omega\bigr]^{2/3}\\
&+\sum_{a,b}\,\sum_{\ell_\star\geq2,\,m_\star}\eta^{\braket{a|b}}_{\ell_\star m_\star}(\Omega)\,e^{-is m_\star\vartheta}\,\bar c_a c_b\,,\\
\eta^{\braket{a|b}}_{\ell_\star m_\star}&\simeq-\frac{4\pi\,q\,\alpha}{2\ell_\star+1}\,\frac{Y_{\ell_\star m_\star}(\tfrac{\pi}{2},0)}{R^{\ell_\star+1}}\int_{\bx}\bar\psi_a\,r^{\ell_\star}Y_{\ell_\star m_\star}(\hat{\bx})\,\psi_b\,,
\end{aligned}
\label{eq:Hbin}
\end{equation}
with $R=[GM(1+q)/\Omega^2]^{1/3}$ the orbital separation and $s=\pm1$ for co- and counter-rotating orbits. This is the leading-order tidal coupling; PN corrections are discussed in~\cite{Boskovic:2026eth}. For a companion outside the cloud, the dipole, $\ell_\star=1$, is absent in the frame of the atom~\cite{Baumann:2018vus,Brito:2023pyl}, so the sum starts with the quadrupole. A term is nonzero only if $m_a=m_b+m_\star$, $|\ell_a-\ell_b|\le\ell_\star\le\ell_a+\ell_b$ and $\ell_a+\ell_b+\ell_\star$ is even~\cite{Baumann:2018vus}. The diagonal terms ($a=b$, $m_\star=0$) shift the levels, while the off-diagonal ones mix them. The mixing of two levels is resonantly enhanced when~\cite{Baumann:2018vus,Baumann:2019ztm}
\begin{equation}
s\,m_\star\Omega^{\mathrm{res}}_{\braket{a|b}} \approx E_a-E_b\,.
\label{eq:res_SM}
\end{equation}
For a free field, Bohr, fine and hyperfine resonances are well separated, at $\Omega/\mu\sim\alpha^2$, $\alpha^4$ and $\tilde a\,\alpha^5$~\cite{Baumann:2018vus,Tomaselli:2024bdd}. Quartic interactions shift them and, for a dense cloud, can move them to a different stage of the inspiral or to the opposite orientation~\cite{Boskovic:2026eth}. For $\{\ket{211},\ket{322}\}$, the resonances met in the early inspiral are all on co-rotating orbits: the hyperfine resonance of $\ket{211}$, $\ket{211}\to\ket{21\,{-1}}$, and the fine resonances of $\ket{322}$, $\ket{322}\to\ket{300}$ ($\ell_\star=2$) and $\ket{322}\to\ket{31\,{\pm1}}$ ($\ell_\star=3$)~\cite{Tomaselli:2024bdd,Boskovic:2025ixx}.

\begin{figure*}
    \centering
    \includegraphics[width=0.99\linewidth]{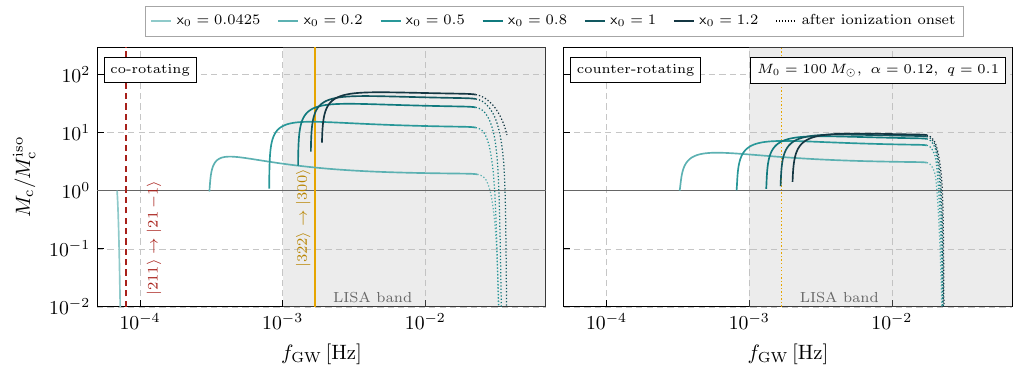}
    \caption{Mass of the cloud with a companion relative to that of an isolated cloud at equal time, $M_{\rm c}/M_{\rm c}^{\rm iso}$, against GW frequency, for the binary of Fig.~\ref{fig:binary} and several formation frequencies $\mathsf{x}_0$ (see text). Curves are solid up to the onset of ionization and dotted after, where they are shown for orientation only. Vertical lines mark the $\ket{322}\to\ket{300}$ resonance (on counter-rotating orbits, where there is no resonance, only its frequency) and the hyperfine resonance of $\ket{211}$.}
    \label{fig:binary_sm}
\end{figure*}

Mixing does not require a resonance, however. Every level has a decay width, so the tide drains a level into a broader one at any orbital frequency, even where Eq.~\eqref{eq:res_SM} cannot be met~\cite{Tomaselli:2025jfo,Boskovic:2025ixx}. The broadest levels are the spherical ones, $\ket{n00}$, with widths $\Gamma_{n00}\sim\mu\alpha^5$, and the mixing $\ket{n\ell m}\to\ket{n00}$ needs $\ell_\star=\ell$. It is therefore absent for $\ket{211}$, which would need the dipole, and quadrupolar for $\ket{322}$. This simplifies the evolution of $\{\ket{211},\ket{322}\}$: away from its hyperfine resonance, $\ket{211}$ is not tidally perturbed, while $\ket{322}$ is adiabatically depleted by the $\ket{322}\to\ket{300}$ mixing (on top of the possible narrow resonances~\cite{Boskovic:2025ixx}), on either orbital orientation and over a large frequency range. This is the off-resonant mixing referred to in the main text.

The mixing backreacts on the orbit through Eq.~\eqref{eq:Hbin}. It induces a small precession of the mean anomaly and changes the orbital frequency at a rate that can compete with GW radiation reaction, and that can either slow or speed up the inspiral, depending on the transition~\cite{Baumann:2019ztm,Boskovic:2024fga,Tomaselli:2024bdd,Tomaselli:2024dbw,Boskovic:2025ixx}. The orbit also responds to the mass that the cloud radiates to infinity, in scalar waves and in GWs. Since $R\,M(1+q)$ is an adiabatic invariant~\cite{Boskovic:2025ixx,us},
\begin{equation}
\frac{\dot\Omega}{\Omega}\supset-\frac{2\,|\dot M_\infty|}{M(1+q)}\,,
\label{eq:Omdot_SM}
\end{equation}
with $\dot M_\infty$ the rate at which mass is radiated. Finally, the companion ionizes the cloud over the same range of frequencies as the Bohr resonances~\cite{Baumann:2021fkf,Tomaselli:2023ysb,Tomaselli:2024bdd}, which can deplete both levels. We do not follow the cloud through ionization and leave this to~\cite{us}.

We evolve the cloud and the orbit together for the binary of Fig.~\ref{fig:binary}, varying the orbital frequency at which the cloud forms, $\mathsf{x}_0\equiv\Omega_0/\Omega^{\mathrm{res}}_{\braket{322|300}}$. Figure~\ref{fig:binary_sm} shows the mass of the cloud relative to that of an isolated cloud at equal time, and Fig.~\ref{fig:binary} shows one of these runs in time. There are two regimes. A cloud that forms below the hyperfine resonance of $\ket{211}$ on a co-rotating orbit is depleted when the orbit crosses it. The SG of the growing cloud pushes this resonance to higher frequencies, so clouds that form somewhat above its free-field position are also partly depleted around saturation. Away from this resonance the tide drains $\ket{322}$ and $\ket{211}$ is preserved, as described in the main text. The gain is largest when the cloud forms close to $\Omega^{\mathrm{res}}_{\braket{322|300}}$ on a co-rotating orbit. At the onset of ionization, $\ket{211}$ is $\mathcal{O}(10)$ times heavier than around an isolated BH on co-rotating orbits, and a factor of a few on counter-rotating ones, where $\ket{322}\to\ket{300}$ is never resonant and the mixing is only off-resonant. In most runs the narrow resonances $\ket{322}\to\ket{31\,{\pm1}}$ are only weakly excited, so we do not track $\ket{31\,{\pm1}}$ occupancies.

\emph{Extending the two-state model.} In the context of Sec.~\ref{app:two_state}, the off-resonant mixing enters Eq.~\eqref{eq:kin} as an extra width of $\ket{b}=\ket{322}$. With $\ket{300}$ integrated out, $\gamma^{\rm SR}_b\to\gamma^{\rm SR}_b-\gamma_{\rm mix}$, where
\begin{equation}
\gamma_{\rm mix}=\Gamma_{300}\,\frac{|\eta|^2}{\Delta^2+\Gamma_{300}^2/4}\,,
\label{eq:gmix}
\end{equation}
with $\eta\equiv\eta^{\braket{322|300}}_{22}$ [Eq.~\eqref{eq:Hbin}],  $\Delta\equiv E_{322}-E_{300}-2s\Omega$ the detuning from the resonance of Eq.~\eqref{eq:res_SM}, and $\Gamma_{300}\simeq0.30\,\tilde r_+\mu\alpha^5$ the width of $\ket{300}$.
We can estimate the strength of this channel in the same way as in \textit{(iii)}, Eq.~\eqref{eq:xend}, finding a tide-generated threshold on the occupancy of $\ket{a}$,
\begin{equation}
\begin{aligned}
x_{\rm tid}&=\left(\frac{\gamma_{\rm mix}}{\gamma_{\rm BH}}\right)^{1/2}\\
&\simeq\frac{q}{1+q}\times
\begin{cases}
460\,\Omega^2/(\mu^2\alpha^7)\,, & \Omega\ll\Omega^{\rm res}_{\braket{322|300}}\,,\\[2pt]
4.4/\tilde r_+\,, & \Omega\simeq\Omega^{\rm res}_{\braket{322|300}} \, (s=+1)\,.
\end{cases}
\end{aligned}
\nonumber
\end{equation}
Since $\ket{211}$ cannot regrow at its threshold, and $\gamma_{\rm mix}$ grows further along the inspiral, the drain of $\ket{211}$ stops at $x_{\rm tid}$ (cf.\ Fig.~\ref{fig:binary}).

Could another level take the place of $\ket{322}$ once the tide has drained it? As noted in~\ref{app:two_state}, the channel $\ket{211}\times\ket{211}\to\ket{n22}\times{\rm BH}$ feeds all overtones $\ket{n22}$, but the tide drains each of them in the same way, through the quadrupolar mixing with $\ket{n00}$. For larger $n$ this mixing is resonant at lower frequencies, $\Omega^{\mathrm{res}}_{\braket{n22|n00}}\propto n^{-3}$, so it is weaker in the range we study. However, the Bohr transitions $\ket{n22}\to\ket{n'00}$, which are similarly broad, and eventually ionization, whose onset scales as $n^{-2}$, move into that range and drain $\ket{n22}$ efficiently. For very large $n$, the overlap of $\ket{n22}$ with $\ket{211}$ decreases, and with it the rate that feeds the overtone. Levels with $\ell_b>2$ are fed only if the absorbed particle has $\ell\geq1$. Such states decay at a rate $\propto\alpha^{4\ell+5}$, far more slowly than those with $\ell=0$, so these channels are inefficient at the small and moderate $\alpha$ considered here. We have checked this in runs that include some of these channels, and $\ket{211}$ survives on intermediate timescales in all of them. A wider exploration will be done in~\cite{us}.

\clearpage 
 
\end{document}